\documentclass[preprint,12pt]{elsarticle}

\usepackage{amssymb}
\usepackage{amsmath}
\usepackage{hyperref}
\usepackage{listings}
\usepackage{xcolor}

\definecolor{keywords}{RGB}{0,125,0}
\definecolor{comments}{RGB}{0,100,113}
\definecolor{red}{RGB}{160,0,0}
\definecolor{green}{RGB}{0,150,0}

\journal{Nuclear Physics A}

\begin{document}

\begin{frontmatter}



\title{Neutron irradiation damage on Silicon Photomultipliers and electrical annealing studies for the CBM RICH detector.}


\author[BUW]{J. Peña-Rodríguez}
\author[BUW]{J. Förtsch}
\author[BUW]{C. Pauly}
\author[BUW]{K.-H. Kampert}

\affiliation[BUW]{organization={Bergische Universität Wuppertal, Fakultät für Mathematik und Naturwissenschaften},
            addressline={Gaußstraße 20}, 
            city={Wuppertal},
            postcode={42119}, 
            state={North Rhine-Westphalia},
            country={Germany}}

\begin{abstract}
Limited radiation hardness is the primary drawback to implementing Silicon Photomultipliers (SiPMs) in high-luminosity environments, such as the Compressed Baryonic Matter (CBM) experiment. Hadron irradiation generates defects in the silicon lattice of SiPMs, increasing dark current, dark count rate (DCR), crosstalk, and afterpulsing, while degrading gain and photon resolution. The expected radiation dose in the photon camera of the Ring Imaging Cherenkov detector of the CBM experiment ranges from $8\times 10^9$ to $5\times 10^{10}$\,n$_{\text{eq}}$/cm$^2$ after two-months of operation at maximum beam energy and intensity. In this work, we evaluated the radiation hardness of three different SiPMs: AFBR-S4N66P024M, S14160-6050HS, and MICROFC-60035. The samples were exposed to neutron irradiation with doses ranging from $3\times 10^8$ to $1\times 10^{11}$\,n$_{\text{eq}}$/cm$^2$. The neutron radiation damage was found to increase the SiPM dark current up to $10^3$ times, DCR up to $10^2$ times, and afterpulsing up to $10\%$ while decreasing their gain and photon resolution. We performed electrical annealing (250\,$^{\circ}$C/30\,min) on the samples to recover the photon resolution and decrease the DCR and dark current.

\end{abstract}

\begin{graphicalabstract}
\includegraphics[width=15cm]{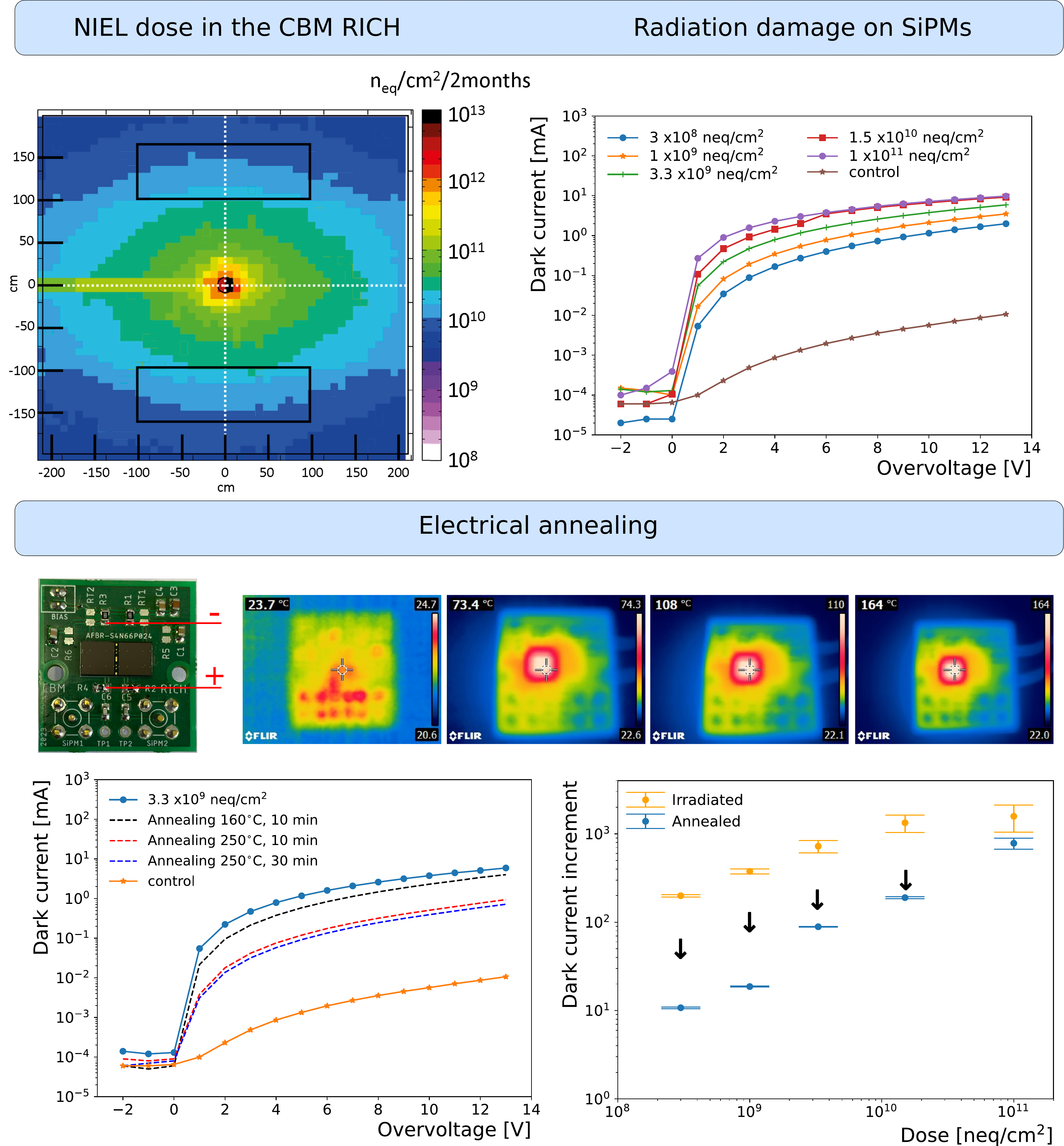}
\end{graphicalabstract}

\begin{highlights}
\item Radiation hardness of three different SiPMs after neutron irradiation from $3\times 10^8$\,n$_{\text{eq}}$/cm$^2$ to $1\times 10^{11}$\,n$_{\text{eq}}$/cm$^2$.
\item Evaluation of in-situ electrical annealing of the irradiated SiPMs. 
\item Assessment of SiPM dark current, DCR, crosstalk, afterpulsing, gain, capacitance, and photon resolution after neutron irradiation and annealing.

\end{highlights}


\end{frontmatter}


\section{Introduction}
\label{intro}

The Compressed Baryonic Matter (CBM) experiment will be located at the future Facility for Antiproton and Ion Research (FAIR) in Darmstadt. The goal of the CBM research program is to explore the QCD phase diagram in the region of high baryon densities using high-energy nucleus-nucleus collisions ($\sqrt{S_{NN}}=2.7 - 4.9$\,GeV). CBM combines to a group of detectors for particle tracking and particle identification (PID). Tracking is achieved by the Micro Vertex Detector (MVD) and the Silicon Tracking Station (STS), while the PID setup chains the Ring Imaging Cherenkov detector (RICH), the Transition Radiation Detector (TRD), and the Time-of-Flight detector (TOF). The CBM RICH detector consists of an ionizing gas chamber (CO$_2$), two semi-spherical mirrors, and two photon cameras. One RICH photon camera forms a matrix of seven rows and 12 columns of back-panels with six H12700 Multi-Anode Photomultipliers (MAPMT) in each back-panel. Each MAPMT contains 64 pixels; the total number of channels in both photon cameras is about $64.5\times 10^3$.  

In recent years, Silicon Photomultiplier (SiPM) technology has been explored as a photodetector candidate for future RICH detectors \cite{Mazziotta2025,Alice2024}. SiPMs have excellent properties in terms of single-photon amplitude discrimination, timing, granularity, and photodetection efficiency (PDE), but low radiation tolerance and high thermal-generated dark noise, especially when aiming for single-photon detection. In nucleus-nucleus collisions, the radiation environment poses an additional challenge. High-energy radiation (protons, neutrons, electrons, and photons) causes surface and bulk damage in the silicon lattice of SiPMs. Radiation damage changes the properties of SiPMs, such as an increase in defects that improve the thermal excitation of charge carriers, a change in the effective doping density that affects the depletion zone width and the electric field, and a decrease in the signal due to the charge trapping effect caused by the lattice defects \cite{Moll_1999}. These processes influence the macro properties of SiPMs due to increasing dark current, dark count rate, and afterpulsing and by decreasing gain, timing, photon resolution, and PDE.

Annealing has been demonstrated to be a reliable methodology for recovering SiPMs operating under high radiation conditions \cite{Gu2023, Preghenella2023, DeAngelis2023, Cordelli2021, Tsang2018}. The process involves heating the silicon lattice above the recrystallization temperature, maintaining a constant temperature for a specified period, and then cooling it down. Annealing reduces radiation damage by recombination after migration of point defects and disassociation of cluster defects \cite{Moll_1999}.

In this paper, we evaluate the radiation damage of SiPMs (AFBR-S4N66P024M, S14160-6050HS, and MICROFC-60035) after neutron irradiation within the dose range expected in the photon cameras of the CBM RICH detector. We performed electrical annealing to evaluate the recovery factor of the irradiated samples. In Section \ref{sec::radiation_damage}, we discuss radiation damage in SiPMs. In Section \ref{sec::radiation_CBM}, we describe the radiation environment of the CBM RICH. In Section \ref{sec::neutron_radiation}, we present the neutron irradiation setup and dark current measurements. In Section \ref{sec::annealing}, we describe the SiPM annealing methodology and the dark current performance at different temperatures and annealing times. Finally, in Section \ref{sec::annealing_results}, we present an evaluation of the SiPM DCR, crosstalk, afterpulsing, gain, capacitance, depletion width, and photon resolution after neutron irradiation.

\section{Radiation damage on SiPMs}
\label{sec::radiation_damage}

High-radiation environments degrade the performance of silicon-based sensors due to surface and bulk damage. Bulk damage is caused by protons, neutrons, $\alpha$ particles, electrons, and photons \cite{Akkerman2001}. When an incoming particle transfers energy to silicon atoms and the energy is higher than the displacement threshold energy ($\sim 25$\,eV), the atoms can be displaced to a new position as interstitial between other atoms. This leaves a vacancy in the crystal lattice. Such defects are called point defects. Single-cluster defects are created at kinetic energies above 1\,keV and multiple-cluster defects for energies above 12\,keV. Electron irradiation mainly causes point defects, while neutron or proton irradiation causes both types of defects \cite{Garutti2019}.

Radiation damage to silicon is proportional to the Non-Ionizing Energy Loss (NIEL) and scales according to the type of particle \cite{Moll_1999}. In photon sensors, radiation damage creates effects such as an increase in the dark current due to defects that facilitate the thermal excitation of electrons/holes, a decrease in signal due to defects that act as trapping centers, and a change in the effective doping density that affects the electric field of the amplification region and the width of the depletion zone \cite{Garutti2019}. 

The leakage current depends on the neutron equivalent fluence $\phi_{eq}$ \cite{Lindstrm2003, Moll_1999},
\begin{equation}
    \Delta I_{dc} = \alpha~ \phi_{eq}~ V ,
\end{equation}
where $V$ is the effective volume and $\alpha$ is the current damage rate. Unlike silicon particle detectors, SiPMs have lower effective volumes, a multiplication region where the electron avalanche is developed, and are composed of thousands of individual avalanche photo-diodes (APDs). Taking these differences into account, we define the SiPM dark current as follows:
\begin{equation}
\label{eq:sipm_idc}
    I_{dc} \sim  \alpha~ \phi_{eq}~ V~ G~ N_p ,
\end{equation}
where $V$ is the effective volume of a single APD, $G$ is the APD gain, and $N_p$ is the number of APDs in the SiPM. The effective volume is,

\begin{equation}
     V \sim S ~d_{\rm eff},
\end{equation}
where $d_{\rm eff}$ is the APD effective thickness (typically 10 - 100\,\textmu m \cite{Riegler2021}) and $S$ the APD area.

\section{Radiation environment at the CBM RICH detector}
\label{sec::radiation_CBM}

\begin{figure}[h!]
    \centering
    \includegraphics[width=0.45\linewidth]{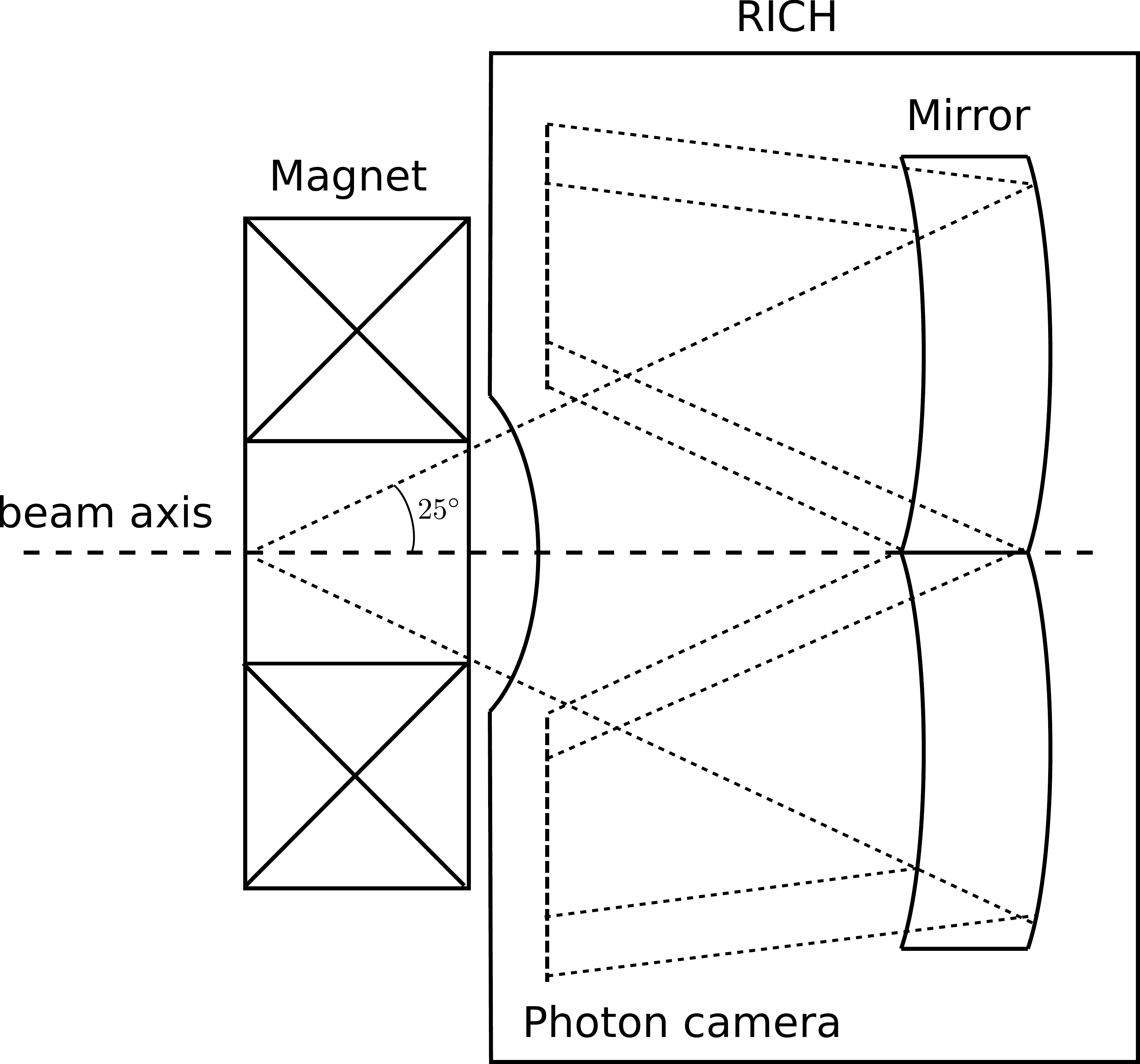}
    \includegraphics[width=0.54\linewidth]{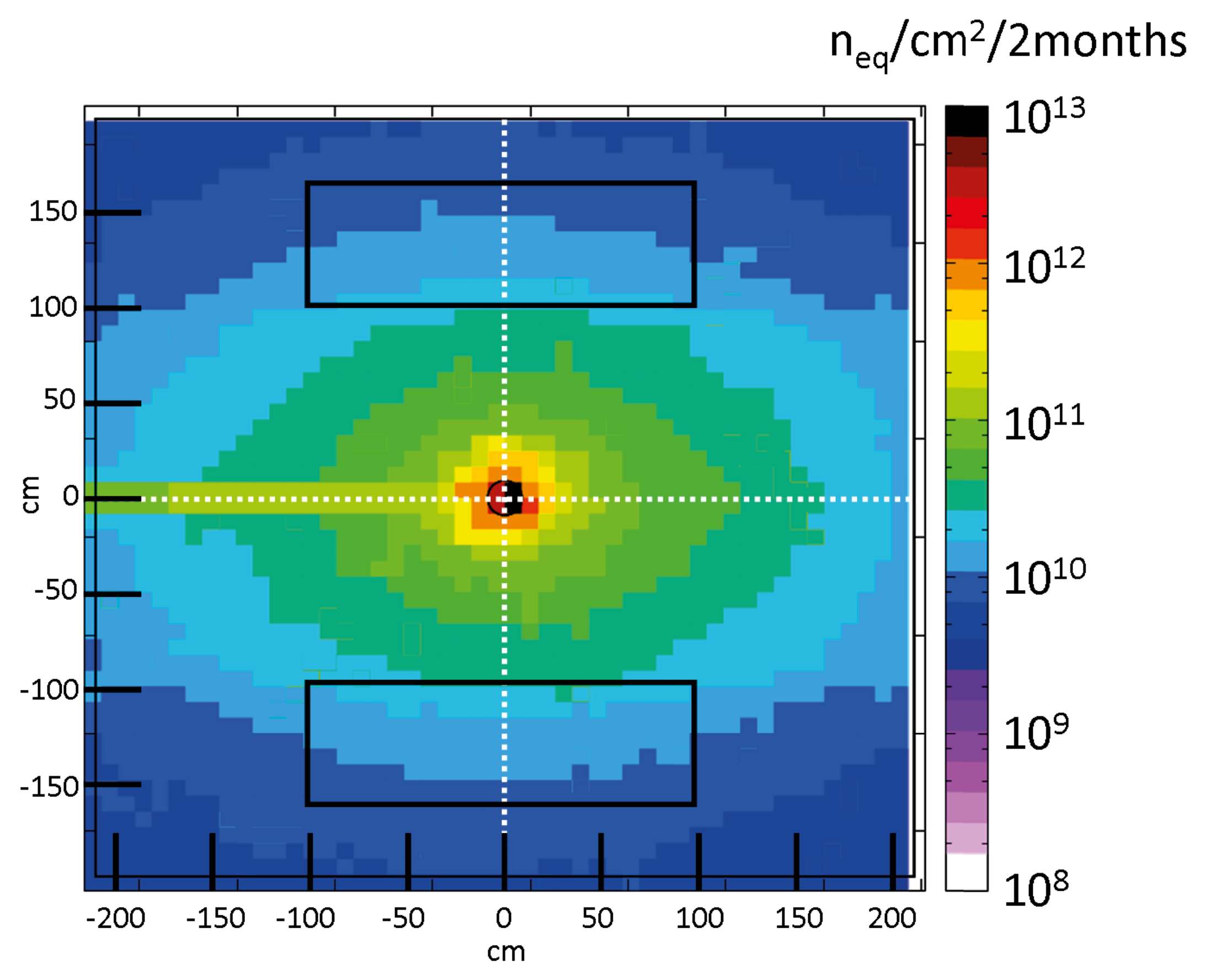}
    \caption{Lateral view of the CBM-RICH detector and magnet (left). Radiation dose of NIEL in CBM-RICH: $8\times 10^9$\,n$_{\text{eq}}$/cm$^2$/2\,months up to $5\times 10^{10}$\,n$_{\text{eq}}$/cm$^2$/2\,months. The black squares represent the photon cameras  (right).}
    \label{fig:NIEL}
\end{figure}

The CBM-RICH photon cameras are exposed to ionizing radiation, neutron flux, and high energy hadron flux, although they are neither placed within the acceptance of the CBM nor directly exposed to the particle flux from the target, as shown in Fig.\,\ref{fig:NIEL}. FLUKA simulations were carried out to estimate the radiation dose on the photon cameras. The results showed doses in ranging from 1\,Gy/2\,months to 20\,Gy/2\,months for ionizing radiation, $8\times 10^9$\,n$_{\text{eq}}$/cm$^2$/2\,months up to $5\times 10^{10}$\, n$_{\text{eq}}$/cm$^2$/2\,months for the NIEL, and $2\times 10^2$\,cm$^{-2}$s$^{-1}$ up to $5\times 10^3$\,cm$^{-2}$s$^{-1}$ for the high-energy hadron flux \cite{sasha2016}. These simulations are a conservative estimate, assuming operation at the largest available beam energy and intensity at FAIR.
Table \ref{tab:rich_dose} shows the radiation doses in all components of the RICH detector: gas vessel, photon detectors, and mirrors. 

\begin{table}[h]
\centering
\scriptsize
\begin{tabular}{ l c c c c c c} \hline
 detector part  & ionization dose & & NIEL & & HE hadron fluence &\\ 
                & [Gy/2 months] & & [n$_{\text{eq}}$/cm$^2$/2 months] & & [1/cm$^2$/s] &\\
                & min & max & min& max & min & max\\  \hline
gas             & 1 & 1000 & 5$\times 10^9$ & 5$\times 10^{12}$ & 100 & 2$\times 10^5$ \\ 
photon detector & 1 & 20 & 8$\times 10^9$ & 5$\times 10^{10}$ & 200 & 5$\times 10^3$ \\ 
mirror          & 2 & 500 & 5$\times 10^9$ & 5$\times 10^{11}$ & 500  & 5$\times 10^4$\\ \hline
\end{tabular}
\caption{Minimum and maximum values of radiation levels (ionization dose, NIEL, and fast hadrons) in the various CBM RICH detector parts, assuming 30~AGeV Au+Au collisions at the maximum anticipated interaction rate of 10~MHz.}
\label{tab:rich_dose}
\end{table}

\section{SiPM neutron irradiation}
\label{sec::neutron_radiation}

We carried out a neutron irradiation campaign of the SiPMs at the U120M cyclotron in the Nuclear Physics Institute CAS, NPI, Department of accelerators located in \v{R}e\v{z}, 25\,km from the center of Prague. The isochronous cyclotron is equipped with beamlines of p, H$^-$, D$^+$, D$^-$, $^3$He$^{2+}$, and $^4$He$^{2+}$.

The neutron beam is created by impinging a proton beam of $\sim$30\,MeV on a 2\,mm thick $^7$Li target followed by a 10\,mm thick carbon absorber to stop the remaining protons. The $^7$Li(p,n) reaction generates a quasi-mono-energetic neutron beam above 20 MeV in the forward direction \cite{Majerle2020}.

\begin{figure}[h!]
    \centering
    \includegraphics[width=0.8\linewidth]{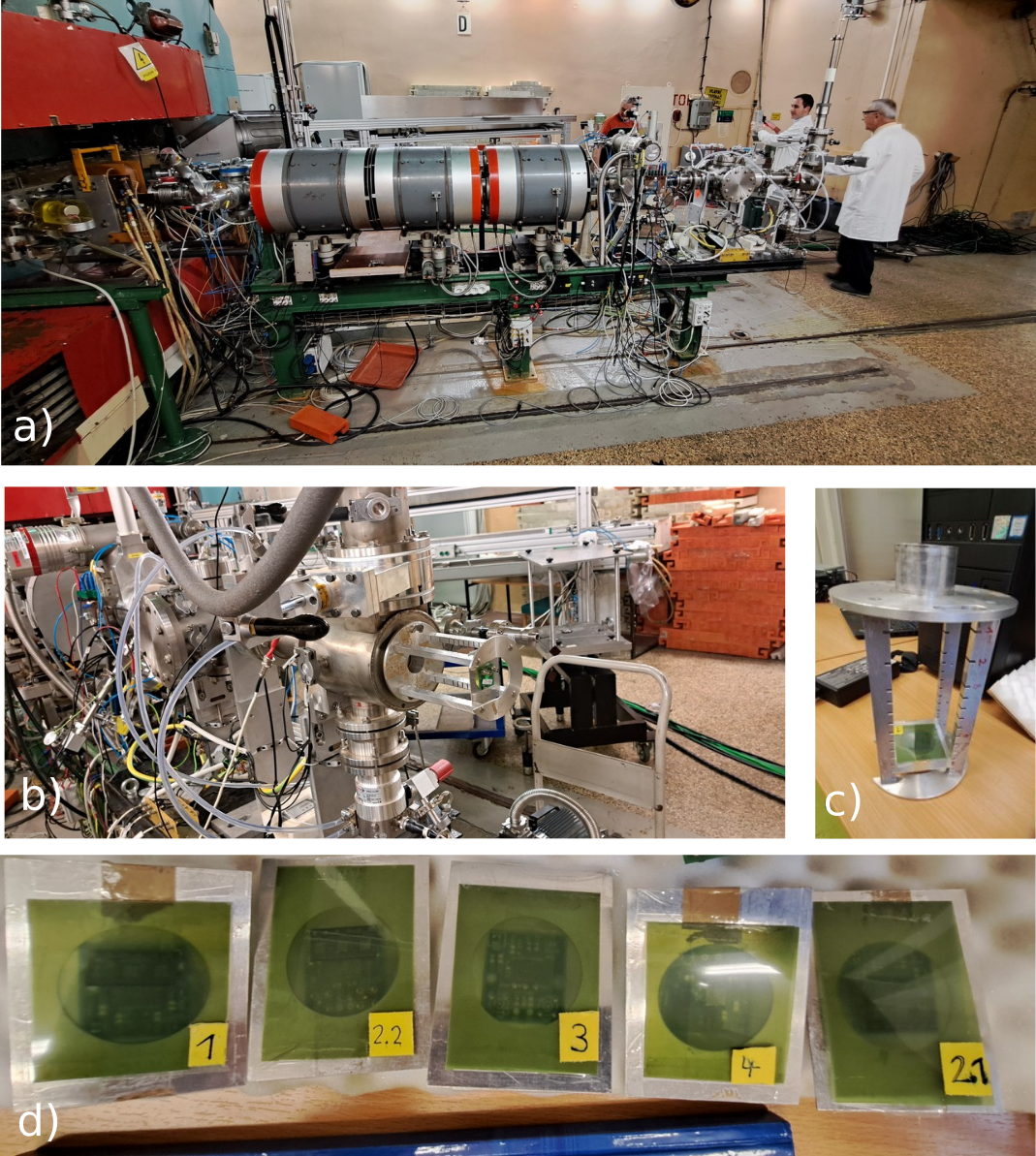}
    \caption{Neutron irradiation setup at the U120M cyclotron. a) Lateral view of the neutron generator. b) Installation of a SiPM sample in the neutron generator. c) SiPM sample holder. d) Preparation of the SiPM samples.}
    \label{fig:beam_setup}
\end{figure}

We irradiated seven SiPM samples: one S14160-6050HS, one MICROFC-60035, and five AFBR-S4N66P024M. The radiation dose ranged from 3.06$\times 10^8$\,n$_{\text{eq}}$/cm$^2$ to 1.08$\times 10^{11}$\,n$_{\text{eq}}$/cm$^2$ and was adjusted by modifying the beam current, the distance of the sample from the production target and the irradiation time, as shown in Table\,\ref{tab:sipm_dose}. The neutron beam setup is shown in Fig.\,\ref{fig:beam_setup}.

\begin{table}[h!]
\centering
\scriptsize
\begin{tabular}{ l c c c c} \\  \hline
sample  & dose [1\,MeV\,n$_{\text{eq}}$/cm$^2$] & beam current [$\%$] & SiPM distance [cm] & irradiation time [s]  \\  \hline
AFBR0                       & 3.06$\times 10^8$ & 10 & 22.6 & 180 \\ 
AFBR1                       & 1.02$\times 10^9$ & 10 & 22.6 & 600 \\ 
AFBR2                       &  &  &  &  \\
S14160                      & 3.28$\times 10^9$ & 10 & 12.6 & 600 \\ 
MICROFC                     &  &  &  &  \\ 
AFBR4                       & 1.53$\times 10^{10}$ & 100 & 22.6 & 900 \\  
AFBR5                       & 1.08$\times 10^{11}$ & 100 & 8.6 & 900 \\  \hline
\end{tabular}
\caption{SiPM radiation dose and neutron beam parameters.}
\label{tab:sipm_dose}
\end{table}

\subsection{Dark current}

\begin{figure}[h!]
    \centering
    \includegraphics[width=0.48\linewidth]{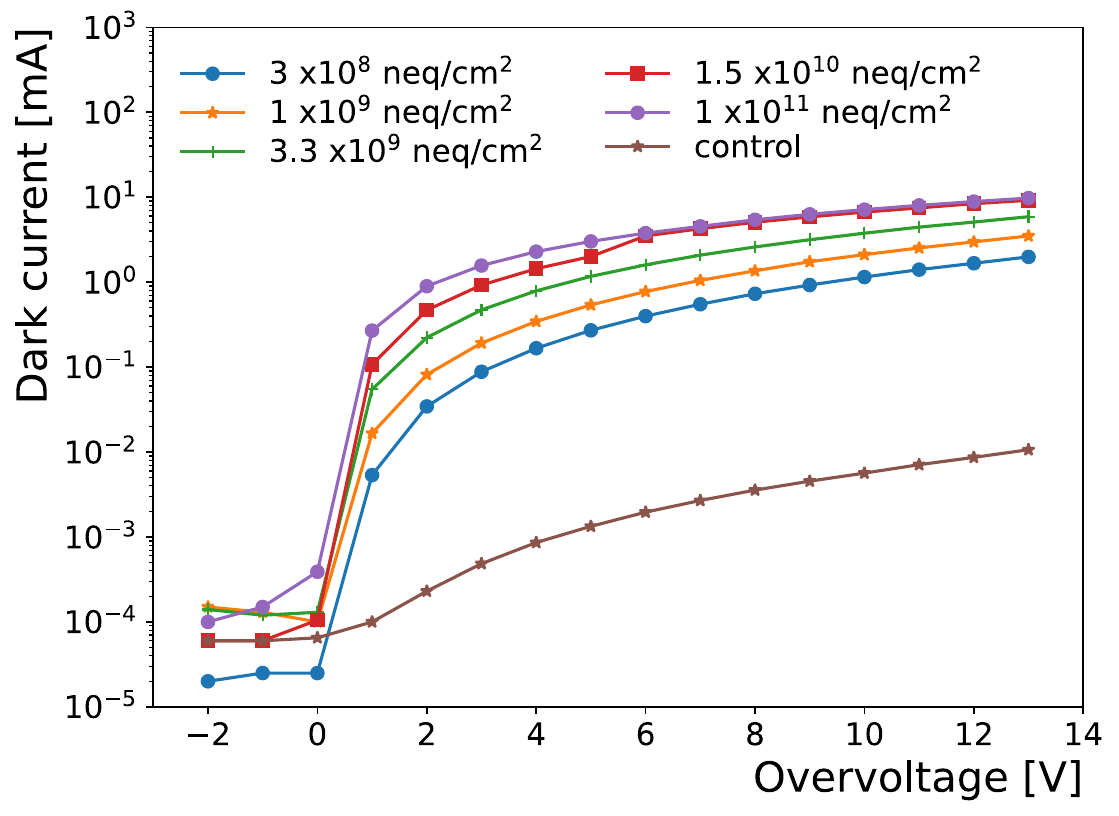}
    \includegraphics[width=0.46\linewidth]{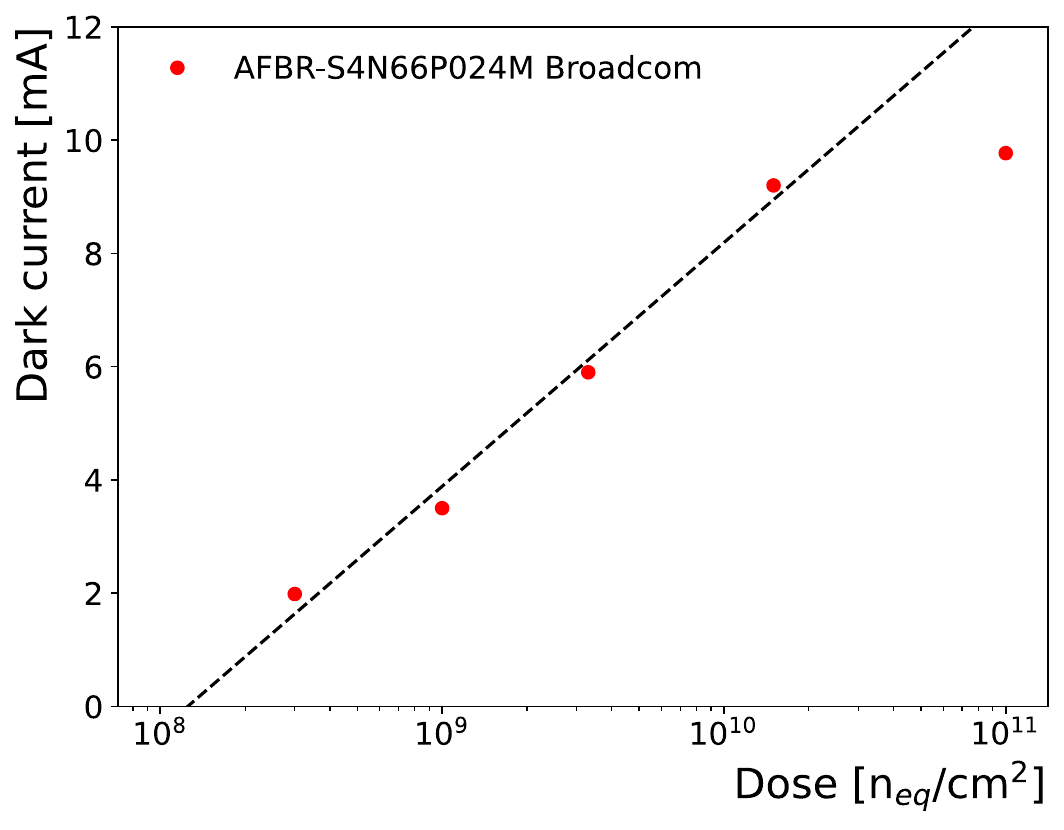}
    \caption{Dark current of the five AFBR-S4N66P024M samples after neutron irradiation (left). Dark current dependency on the neutron radiation dose (right).}
    \label{fig:darkcurrent}
\end{figure}

We measured the dark current of the irradiated samples after a 2-month storage period at room temperature. Each SiPM sample was mounted inside a light-tight box connected to an ammeter and a power supply in series. Figure\,\ref{fig:darkcurrent} shows the measured dark current of the SiPM samples. In the left panel, the dark current of the five AFBR-S4N66P024M samples at different radiation doses ($3 \times 10^8$\,n$_{\text{eq}}$/cm$^2$, $1 \times 10^9$\,n$_{\text{eq}}$/cm$^2$, $3.3 \times 10^9$\,n$_{\text{eq}}$/cm$^2$, $1.5 \times 10^{10}$\,n$_{\text{eq}}$/cm$^2$, and $1 \times 10^{11}$\,n$_{\text{eq}}$/cm$^2$) is shown. The dark current ($I_{dc}$) of the SiPM increases linearly with radiation dose. The right panel of Fig.\,\ref{fig:darkcurrent} shows the dark current of the SiPMs as a function of the radiation dose for an operation bias of 43\,V/20\,$^{\circ}$C.

We defined the dark current ratio to assess the dark current increment,

\begin{equation}
    R = \frac{\text{irradiated} \ I_{dc}}{\text{control} \ I_{dc}} \: .
\end{equation}

\begin{figure}[h!]
    \centering
    \includegraphics[width=0.6\linewidth]{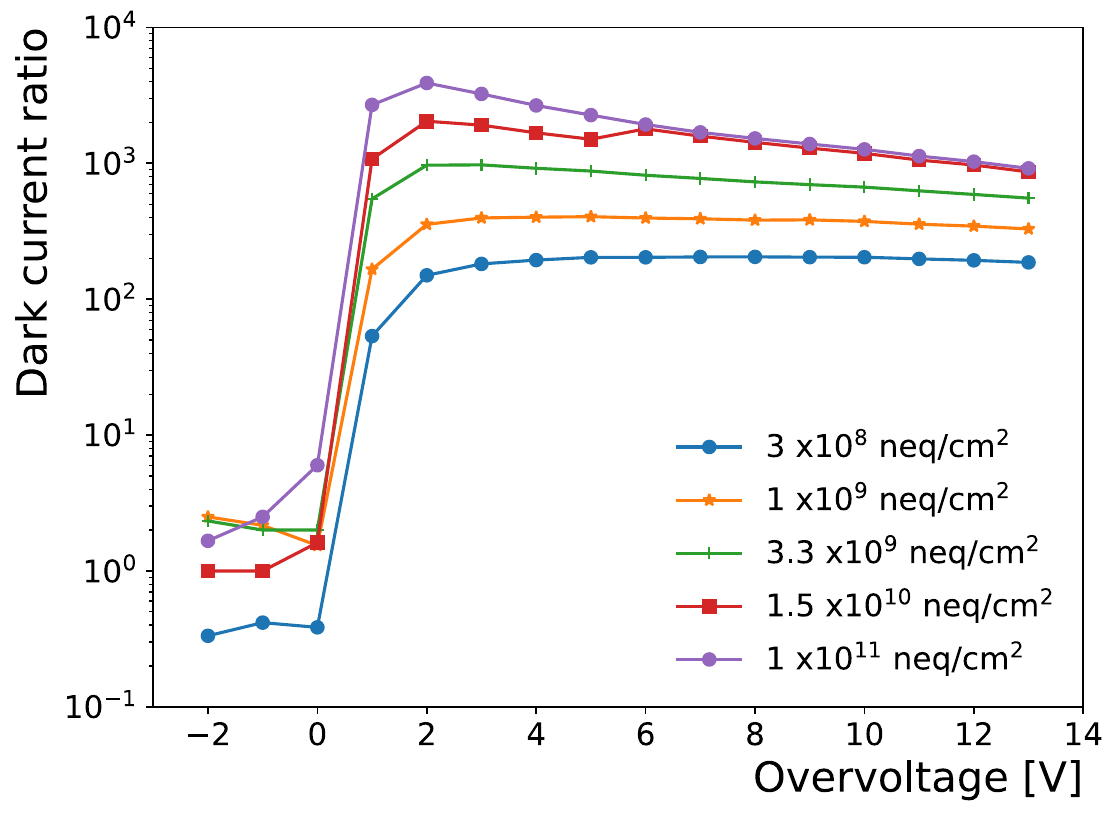}
    \caption{Dark current ratio of the AFBR-S4N66P024M samples after neutron irradiation  operating at overvoltages ranging from 0 to 13\,V. }
    \label{fig:darkcurrent_ratio}
\end{figure}

As shown in Fig.\,\ref{fig:darkcurrent_ratio}, the dark current of the sample irradiated at $3 \times 10^8$\,n$_{\text{eq}}$/cm$^2$ increased by a factor of $\sim 2 \times 10^2$ while the dark current of the sample irradiated at $1 \times 10^{11}$\,n$_{\text{eq}}$/cm$^2$ increased by a factor of $\sim 2\times 10^3$. The increase in dark current is primarily due to the creation of point and cluster defects in the silicon lattice resulting from the neutron beam.

\begin{figure}[h!]
    \centering
    \includegraphics[width=0.6\linewidth]{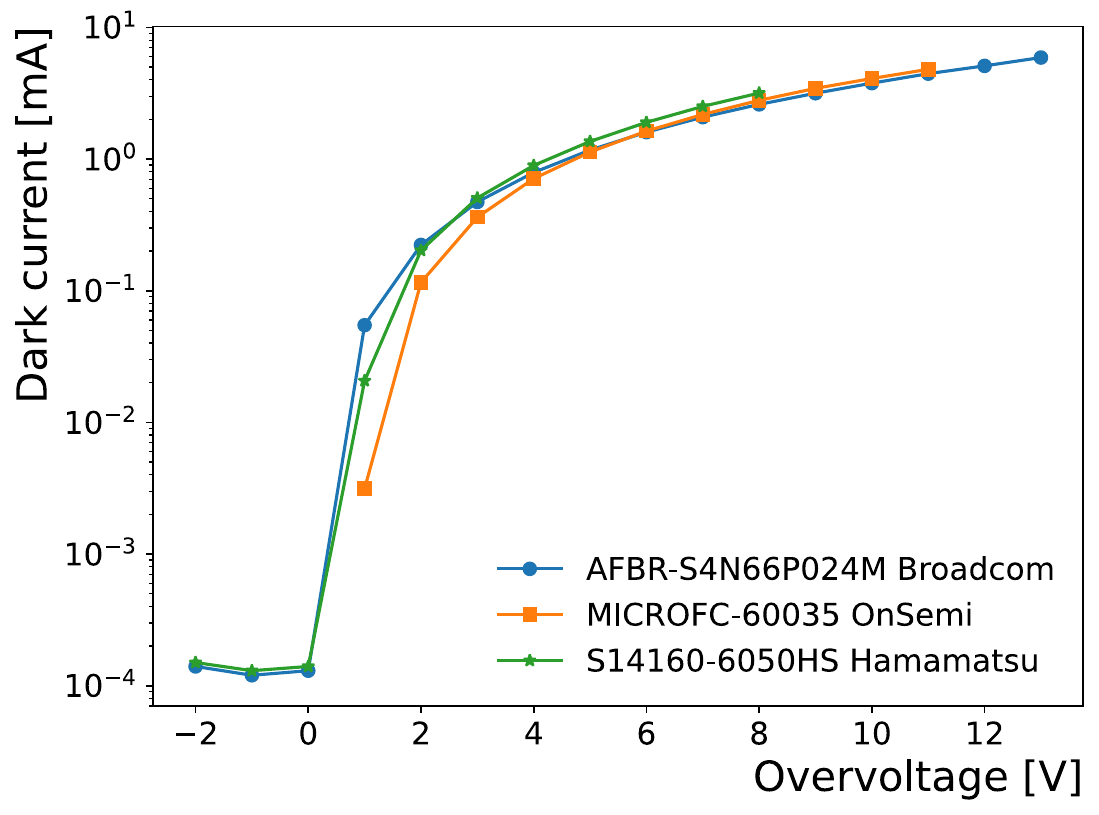}
    \caption{Dark current depending on over-voltage for the AFBR-S4N66P024M, S14160-6050HS, and MICROFC-60035 SiPMs. }
    \label{fig:3sipm_darkcurrent}
\end{figure}

Depending on the inner structure of the SiPM, the dark current could vary. We evaluated the radiation damage of three different SiPMs. Figure\,\ref{fig:3sipm_darkcurrent} shows the dark current of the AFBR-S4N66P024M, S14160-6050HS, and MICROFC-60035 SiPMs after irradiation at $3.3 \times 10^9$\,n$_{\text{eq}}$/cm$^2$. We observed that the dark current behaves similarly for the three SiPMs at a given overvoltage value, but it is slightly lower for the MICROFC-60035. As shown in Equation \ref{eq:sipm_idc}, the dark current depends on the radiation dose, the silicon volume, the gain, and the number of pixels. Table\,\ref{tab::sipm} shows the parameters of the studied SiPMs.

\begin{table}[h]
    \centering
    \begin{tabular}{lccc}\hline
        SiPM & MICROFJ-60035 & S14160-6050CS  & AFBRS4N66P024\\
         & OnSemi & Hamamatsu & Broadcom \\ \hline
        Pixel pitch (\textmu m) & 35 & 50  & 40\\
        N. of pixels & 22292 & 14331 & 22428\\
        $^\dagger$Gain ($\times10^6$) & $\sim$4.7 & $\sim$3.8 & $\sim$3.2\\  \hline
    \end{tabular}
    \caption{Parameters of the MICROFJ-60035, S14160-6050CS, and AFBRS4N66P024. \footnotesize$^\dagger$Gain at 4\,V overvoltage.}
    \label{tab::sipm}
\end{table}

\section{Electrical annealing}
\label{sec::annealing}

Studies have shown that annealing techniques are effective in mitigating radiation damage in SiPMs \cite{Gu2023, Preghenella2023, DeAngelis2023, Cordelli2021, Tsang2018}. Annealing can be carried using microwave, laser, hot plate, storage, or electrical methods. Electrical annealing involves heating the SiPM through the Joule effect by applying a reverse or forward current to the device. Unlike storage or hot plate methods, electrical annealing takes minutes or hours instead of days. Another advantage of electrical annealing is that the annealing setup can be embedded in the SiPM electronics front-end, because of its electrical nature. In summary, electrical annealing enables a fast in-situ recovery of SiPMs operating in high-radiation environments. 

\begin{figure}[h!]
    \centering
    \includegraphics[width=1\linewidth]{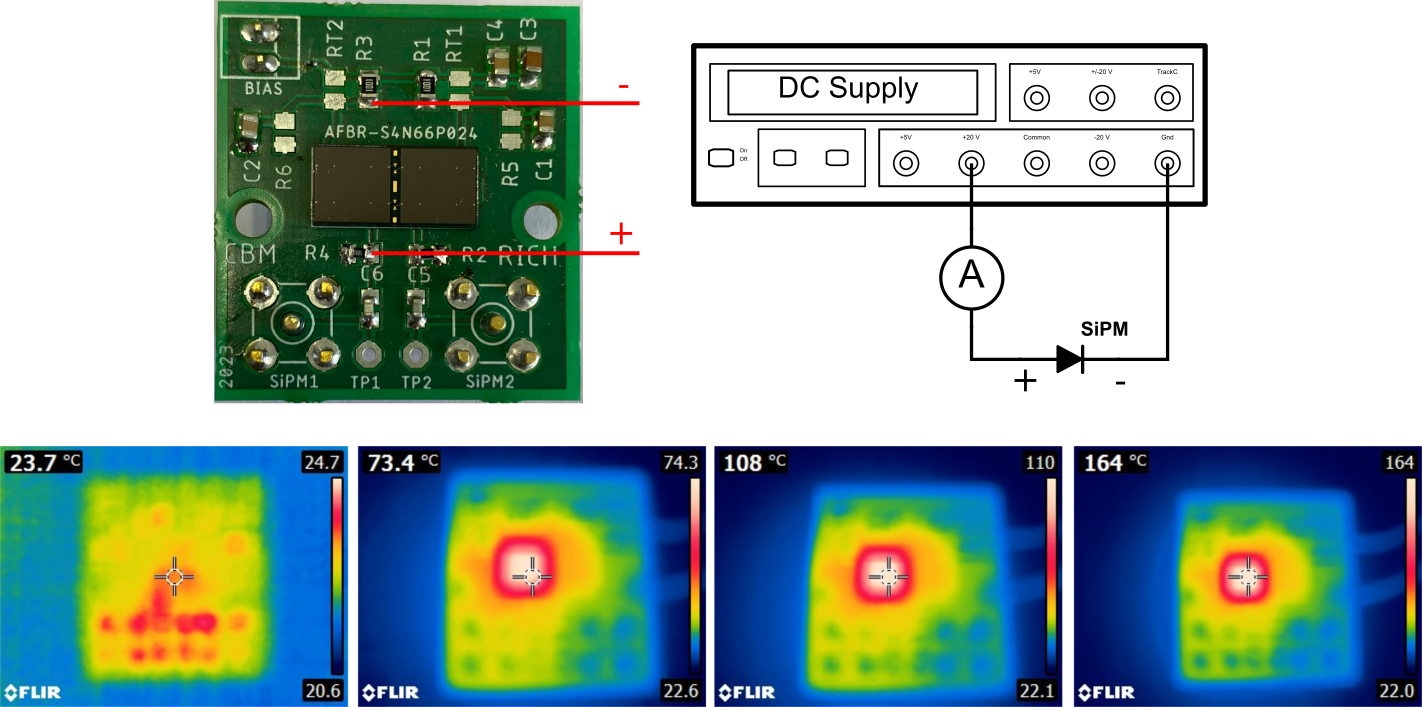}
    \caption{Electrical annealing setup. Forward polarization of the AFBR-S4N66P024M SiPM and annealing setup (top). Thermal camera histograms of the SiPM PCB during the heating process at 23\,$^{\circ}$C (room temperature), 73\,$^{\circ}$C, 108\,$^{\circ}$C, and 164\,$^{\circ}$C respectively (bottom). }
    \label{fig:annealing_setup}
\end{figure}

We performed electrical annealing on the neutron-irradiated samples. Figure\,\ref{fig:annealing_setup} shows the annealing setup. The SiPM samples were forward polarized (1-10\,V) while the current and temperature (FLIR E6 thermal camera) were measured. 

\begin{figure}[t]
    \centering
    \includegraphics[width=0.48\linewidth]{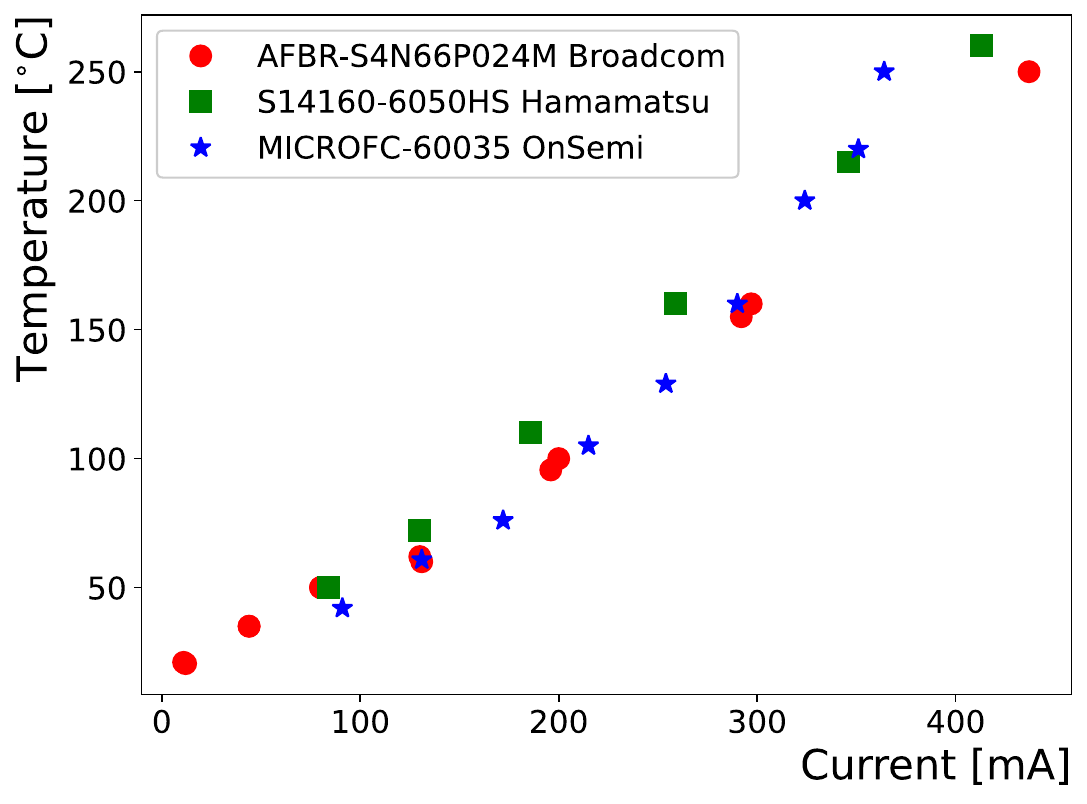}
    \includegraphics[width=0.48\linewidth]{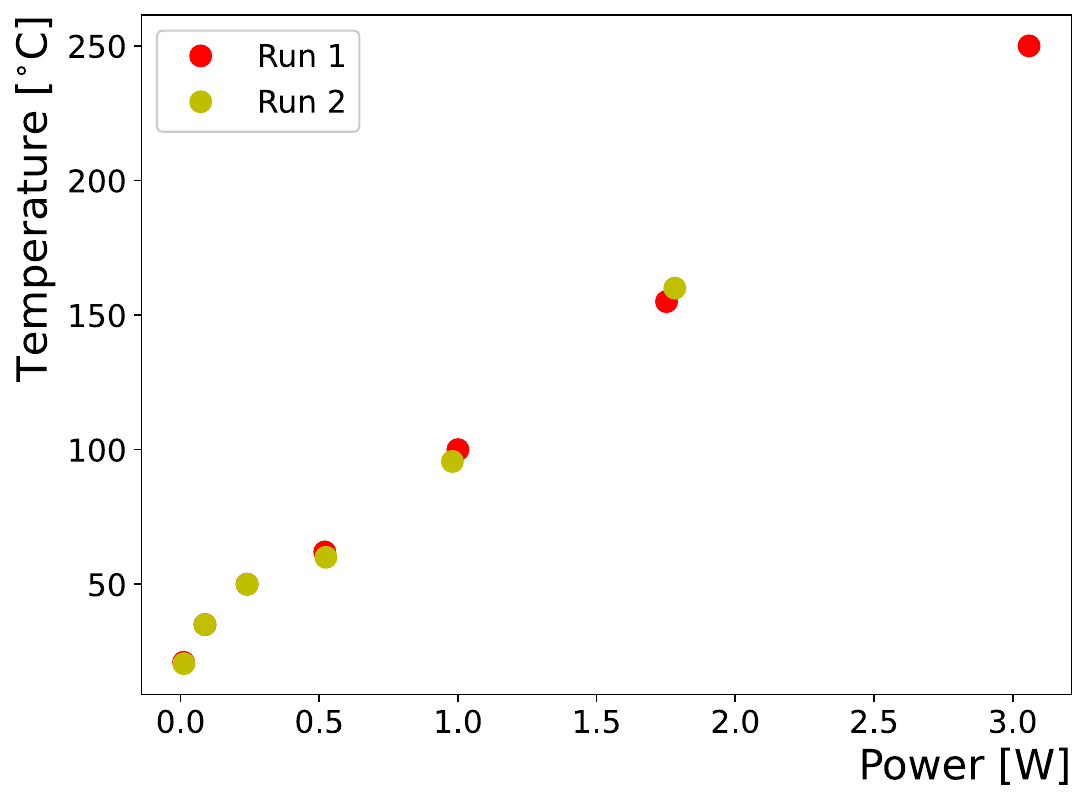}
    \caption{Annealing temperature depending on the SiPM forward current for the AFBR-S4N66P024M (red), S14160-6050HS (green), and MICROFC-60035 (blue) SiPMs (left). Electrical power of the AFBR-S4N66P024M sample during two test runs (right).}
    \label{fig:annealing_temp}
\end{figure}

Figure\,\ref{fig:annealing_temp} shows the annealing temperature as a function of the forward current for the AFBR-S4N66P024M, S14160-6050HS, and MICROFC-60035 SiPMs. The annealing temperature increases nearly linearly for the AFBR-S4N66P024M and S14160-6050HS, but the MICROFC-60035 exhibits a different, more non-linear behavior, particularly above 150\,$^{\circ}$C. This different temperature behavior is probably caused by a different PCB geometry: The MicroFC SiPM formed a double-stack of two PCBs, whereas the other two SiPMs were soldered directly onto the measurement PCB. The annealing temperature characteristic allows us to set a specific SiPM temperature by applying a specific forward current during the in-situ annealing process. The power consumption of the annealed samples reaches about 3\,W at 250\,$^{\circ}$C.

We evaluated the exposure time and temperature of the SiPM annealing process. The annealing profile starts at room temperature (20\,$^{\circ}$C), increases to the annealing temperature (160\,$^{\circ}$C or 250\,$^{\circ}$C) over approx.\ 1\,min, and then holds the temperature constant for 10-30\,minutes. Finally, the SiPM temperature decreases back to room temperature over 5 minutes.

\begin{figure}[th!]
    \centering
    \includegraphics[width=0.48\linewidth]{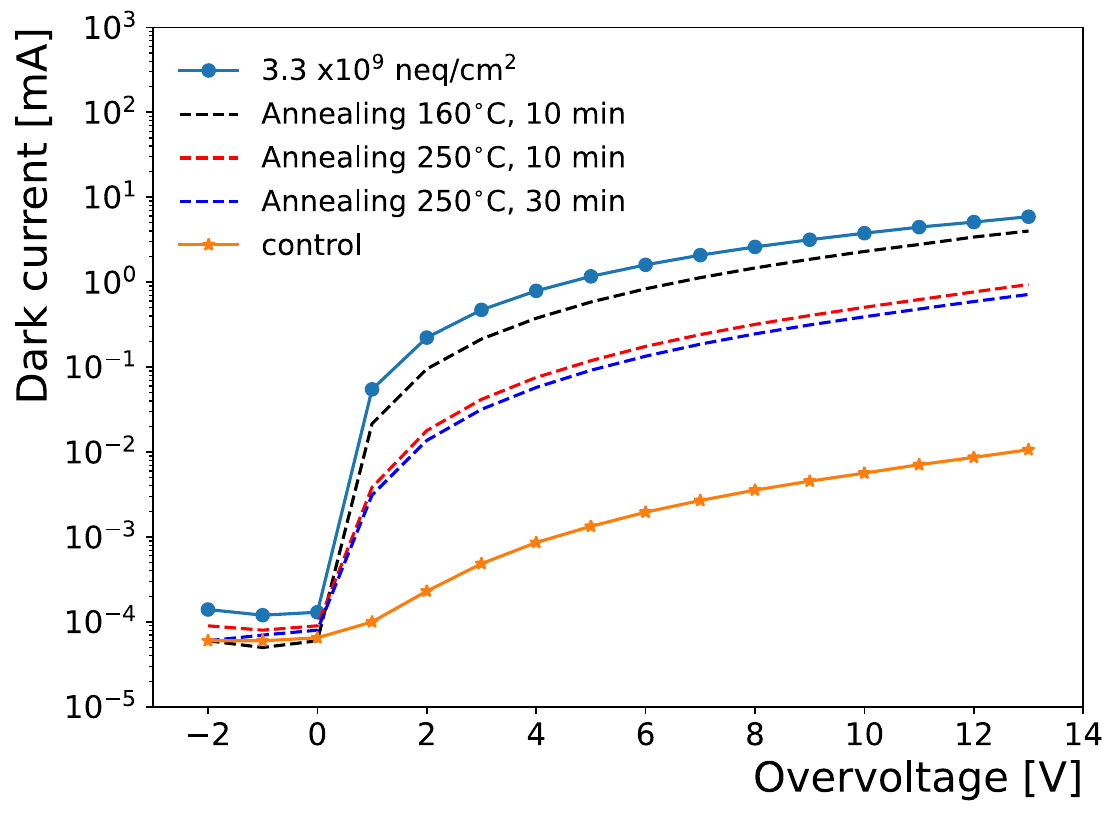}
    \includegraphics[width=0.48\linewidth]{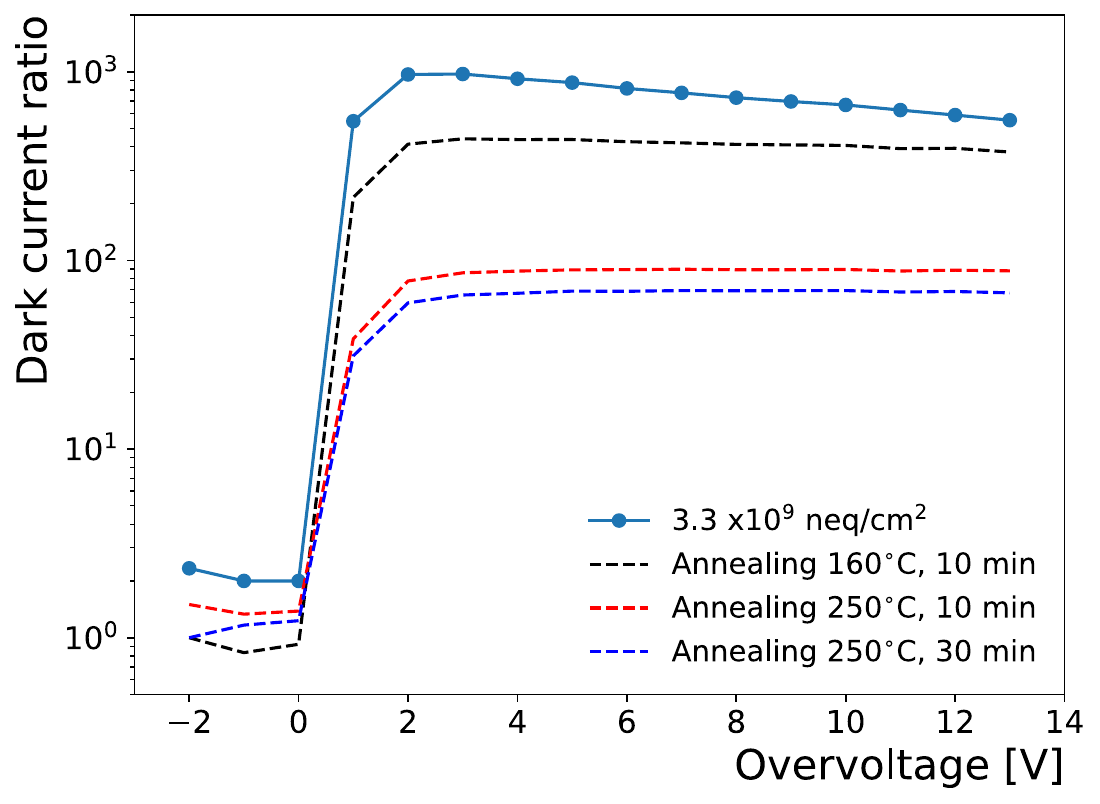}
    \caption{Dark current of the AFBR-S4N66P024M SiPM irradiated at $3.3 \times 10^9$\,n$_{\text{eq}}$/cm$^2$ after annealing at 160\,$^{\circ}$C during 10\,min and at 250\,$^{\circ}$C during 10\,min and 30\,min (left). Dark current increasing factor relative to the control sample (right).}
    \label{fig:annealing_afbr2a}
%
    \vspace{5mm}\centering
    \includegraphics[width=0.48\linewidth]{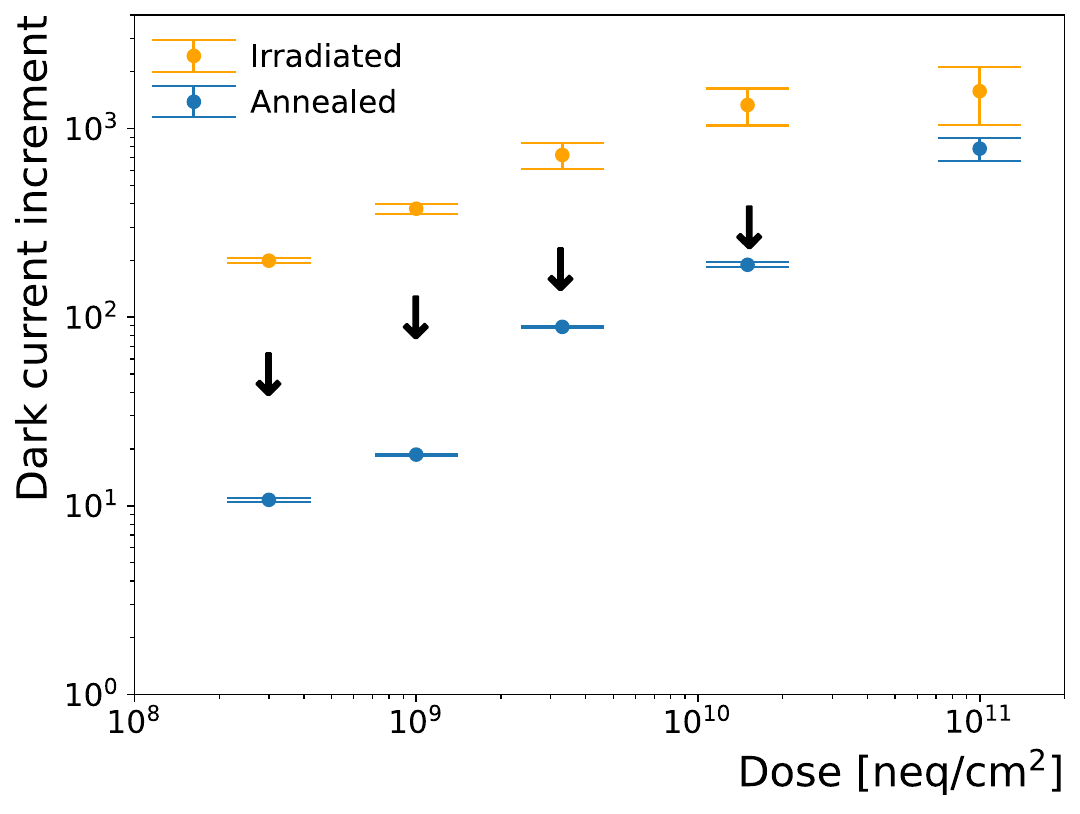}
    \includegraphics[width=0.48\linewidth]{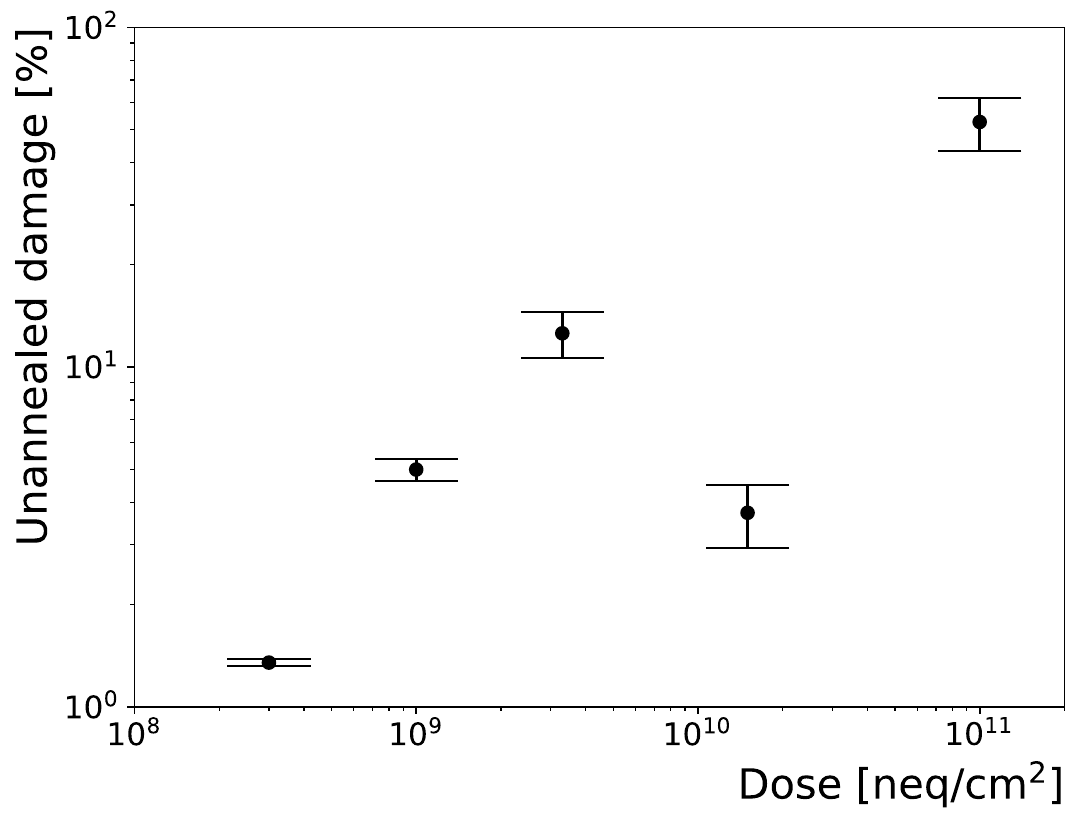}
    \caption{Dark current ratio of the AFBR-S4N66P024M samples after irradiation ($3 \times 10^8$\,n$_{eq}$/cm$^2$, $1 \times 10^9$\,n$_{\text{eq}}$/cm$^2$, $3.3 \times 10^9$\,n$_{\text{eq}}$/cm$^2$, $1.5 \times 10^{10}$\,n$_{\text{eq}}$/cm$^2$, and $1 \times 10^{11}$\,n$_{\text{eq}}$/cm$^2$) and annealing (250\,$^{\circ}$C/30\,min)(left). Unannealed damage of the irradiated AFBR-S4N66P024M samples (right).}
    \label{fig:annealing_results}
\end{figure}

Figure\,\ref{fig:annealing_afbr2a} shows the dark current of the AFBR-S4N66P024M after being irradiated at $3.3 \times 10^9$\,n$_{\text{eq}}$/cm$^2$ and then annealed at 160\,$^{\circ}$C for 10\,min, and at 250\,$^{\circ}$C for 10\,min and 30\,min, respectively. In the first case, the dark current decreased by a factor of $1.9\pm0.3$. At 250\,$^{\circ}$C, the current decreased by a factor of $9.2\pm2.3$ ($12\pm2.8$) after 10 (30)\,min, respectively. 

We observed that the dark current decreases with the annealing time until it reaches a saturation region, where no further recovery is possible. Although annealing can undo vacancies, interstitials, and dislocations in the silicon lattice caused by irradiation, deep-level traps and structural changes are irreversible. This irreversible effect is called unannealed damage,
\begin{equation}
    R = (I_{\phi, T}~/~I_{\phi})~\times~100\% ,
\end{equation}
where $I_{\phi}$ denotes the dark current after irradiation with flux $\phi$, $I_{\phi, T}$ denotes the dark current after irradiation and annealed at temperature $T$. Figure\,\ref{fig:annealing_results} (left) shows the dark current of the AFBR-S4N66P024M irradiated samples (yellow) and annealed (blue) at 250\,$^{\circ}$C during 30\,min. 

The absolute recovery of the irradiated sample depends on the radiation dose. The greater the dose, the less the reduction in dark current reduction with annealing. The sample irradiated at $3 \times 10^8$\,n$_{\text{eq}}$/cm$^2$ experienced a dark current increase of $\sim 2 \times 10^2$, which decreased to $\sim 1 \times 10^1$ after annealing, resulting in an unannealed damage of 1.35\,\%. Conversely, the sample that was irradiated at $1 \times 10^{11}$\,n$_{\text{eq}}$/cm$^2$ had a dark current increase of $\sim 1.58\times 10^3$, which decreased to $\sim 7.8 \times 10^2$, resulting in an unannealed damage of 52.6\,\% as shown in Fig.\,\ref{fig:annealing_results} (right).

\section{SiPM performance after annealing}
\label{sec::annealing_results}

The evaluation of the SiPM performance after irradiation and annealing involved analyzing the DCR, crosstalk, afterpulsing, gain, and photon resolution.

Quantifying the noise of SiPMs after irradiation helps design robust systems for particle detectors that operate at high radiation doses. Several studies have analyzed the noise of SiPMs after irradiation. Garutti et al.\ present a method for determining the dark noise and afterpulses of KETEK SiPMs after neutron radiation of 1-10$\times10^{10}$\,n$_{\text{eq}}$/cm$^2$. They observed a 3-order-of-magnitude increase in noise, but could not disentangle the afterpulse from the dark noise due to the high noise level \cite{Garutti2014}. Altamura et al.\ estimated the SiPM correlated noise after proton irradiation reaching fluences of up to $10^{14}$\,n$_{\text{eq}}$/cm$^2$ \cite{Altamura2022}.  Correlated noise is the sum of crosstalk, delayed crosstalk, and afterpulses. They concluded that DCR events occurring very close together could be mistaken for crosstalk events. This effect increases with the DCR and decreases with increasing time resolution of the acquisition system. Xu et al.\ demonstrated that SiPM DCR and crosstalk escalate following X-ray irradiation at doses of 200\,Gy, 20\,kGy, 2\,MGy, and 20\,MGy \cite{Xu2014}. Ulyanov et al.\ conducted noise studies after proton irradiation ranging from 1.27$\times10^{8}$\,n$_{\text{eq}}$/cm$^2$ to 1.23$\times10^{10}$\,n$_{\text{eq}}$/cm$^2$, observing an increase in dark noise. Dark noise was defined as the sum of DCR, crosstalk, and afterpulsing \cite{Ulyanov2020}.

\subsection{Dark count rate (DCR) and dark current}

The dark current of a SiPM pixel (Avalanche Photo Diode, APD) can be modeled as
\begin{equation}
    I_{dc} = e\times G \times \text{DCR} \: ,
\end{equation}
where $e$ is the electron charge, $G$ is the APD gain, and DCR is the APD's dark count rate \cite{Pagano2012}. 

A SiPM consists of thousands of APDs connected in parallel, forming a matrix of neighboring pixels. This configuration facilitates the generation of crosstalk, introducing a new component to the dark current estimation. Crosstalk occurs when photons emitted during the primary avalanche are absorbed by neighboring APDs surrounding the primary APD. The crosstalk is estimated as the ratio of events above 1.5\,pe to events above 0.5\,pe. However, this ratio obscures information about the released charge, depending on the number of secondary avalanches. One crosstalk event can trigger multiple secondary avalanches, which breaks the linearity between the dark current and the DCR.
%
%
Afterpulsing also introduces nonlinearities into the expression because the charge of an afterpulse is lower than that of a dark count event, depending on the trapping and recovery times \cite{PeaRodrguez2025}. 

We estimated the DCR for the AFBR-S4N66P024M samples after irradiating them at $3 \times 10^8$\,n$_{\text{eq}}$/cm$^2$, $1 \times 10^9$\,n$_{\text{eq}}$/cm$^2$, and $3.3 \times 10^9$\,n$_{\text{eq}}$/cm$^2$ and annealing them at 250\,$^{\circ}$C for 30\,min. Samples irradiated at the highest doses ($1.5 \times 10^{10}$\,n$_{\text{eq}}$/cm$^2$ and $1 \times 10^{11}$\,n$_{\text{eq}}$/cm$^2$) were excluded because the estimation of the DCR becomes highly difficult due to the pileup effect.

\begin{figure}[h!]
    \centering
    \includegraphics[width=0.6\linewidth]{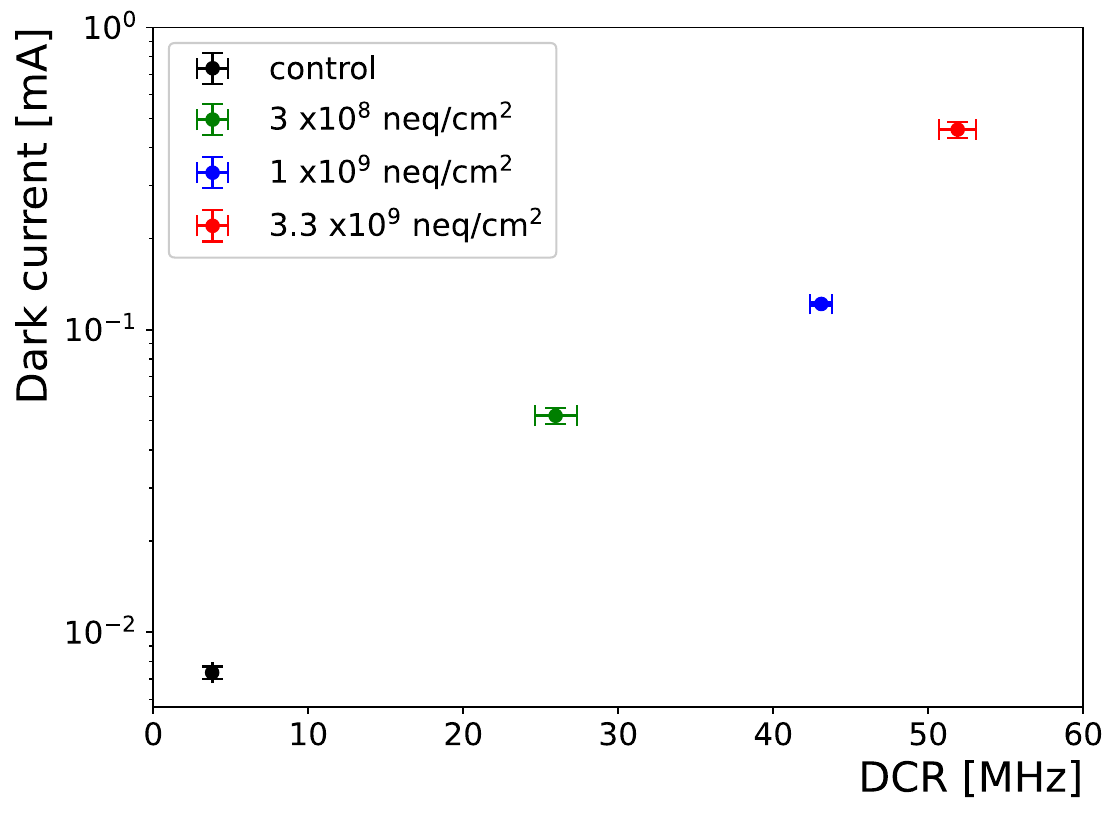}
    \caption{Dark current dependency on DCR after irradiation and annealing of AFBR-S4N66P024M samples operating at 43\,V.}
    \label{fig:Idc_DCR}
\end{figure}

Figure\,\ref{fig:Idc_DCR} shows the relationship between the dark current and the DCR of the AFBR-S4N66P024M samples after irradiation and annealing. The dark current increases exponentially with respect to the DCR.

\subsection{Crosstalk and afterpulsing}

\begin{figure}[t]
    \centering
    \includegraphics[width=0.7\linewidth]{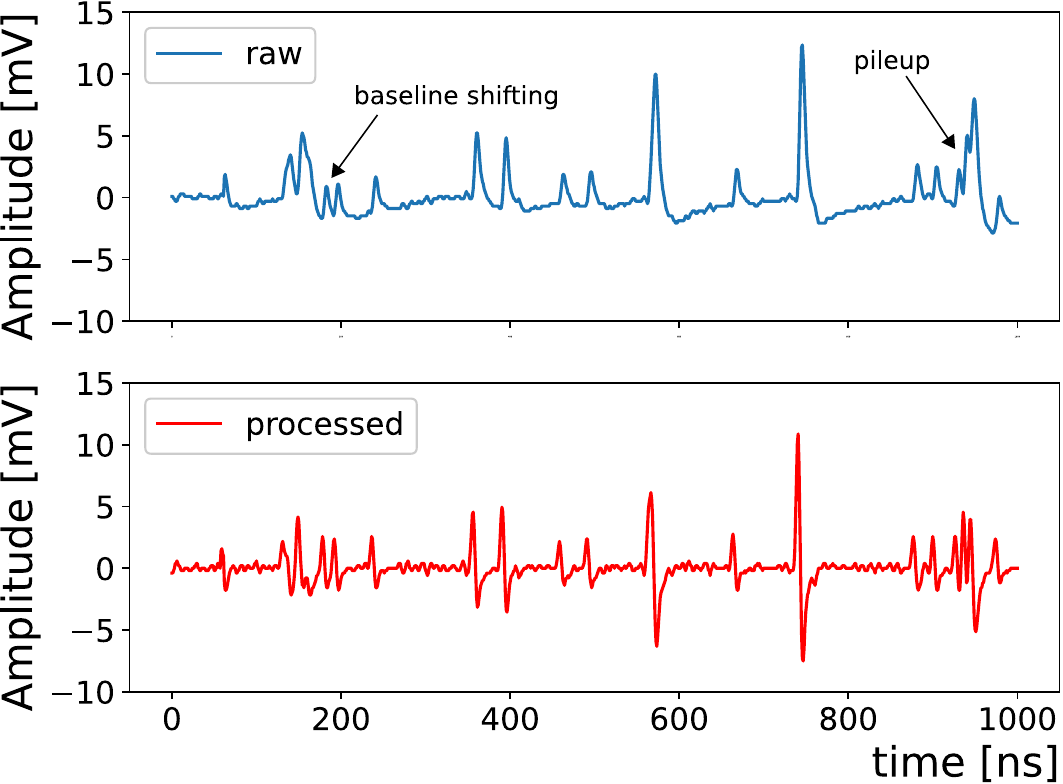}
    \caption{Signal processing before noise analysis. Raw SiPM signal (top) and processed signal (bottom).}
    \label{fig:derivate}
\end{figure}

We carried out a novel analysis technique to isolate the correlated noise components after irradiation. We extracted the amplitude and the time interval between consecutive pulses from the SiPM and removed the baseline shift caused by the pulse undershoot after signal amplification and pulse pileup by differentiating the sampled signal stream. Figure\,\ref{fig:derivate} (top) illustrates the impact of undershoot and pileup on the SiPM signal. The bottom shows how the differentiating filter mitigates these effects.

\begin{figure}[h!]
    \centering
    \includegraphics[width=0.7\linewidth]{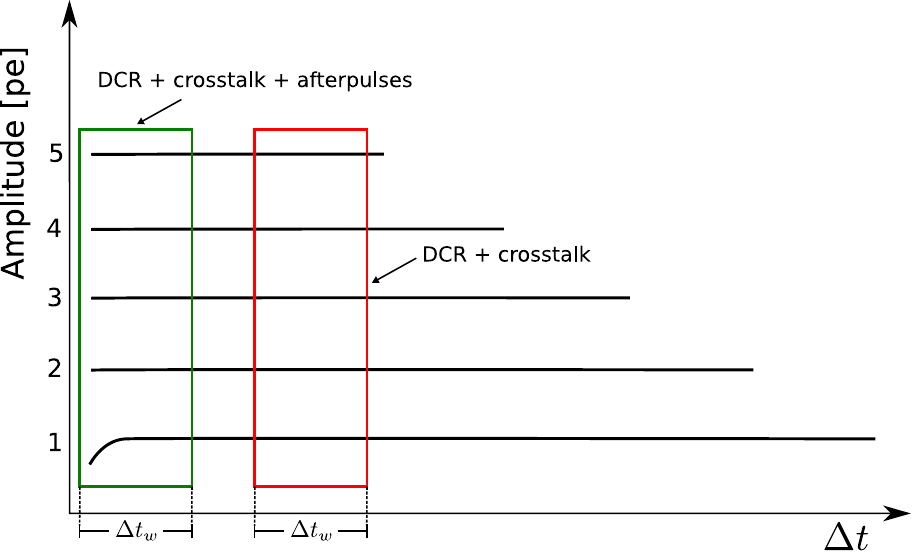}
    \caption{Windowing of the amplitude vs.\ relative time spectrum for estimating DCR, crosstalk, and afterpulsing. Crosstalk is estimated as the ratio between the number of events above 1.5\,pe and the number of events above 0.5\,pe within the red-box. Afterpulsing results from the subtraction of the DCR and crosstalk components from the green-box.}
    \label{fig:noise}
\end{figure}

We estimate DCR, crosstalk, and afterpulsing using a two-dimensional representation of signal amplitude and time interval between consecutive events. We divide the spectrum into two regions to isolate noise components. Figure\,\ref{fig:noise} shows an example spectrum with maximum crosstalk of 5\,pe. The red box includes DCR and crosstalk; the green box includes DCR, crosstalk, and afterpulsing. In regions with only DCR and crosstalk, we can accurately estimate the crosstalk ratio. Afterpulsing is then estimated by subtracting DCR and crosstalk components from the events contained in the green box.


We defined the correlated noise ratio (CNR) as the ratio between events above 1.5\,pe ($N_{\text{1.5pe}}$) and events above baseline noise ($N_{BLN}$) for any window width $\Delta t_w = \Delta t_2 - \Delta t_1$, where $\Delta t_2 > \Delta t_1$. 

\begin{equation}
    \text{CNR} = \frac{N_{\text{1.5pe}}}{N_{BLN}}.
\end{equation}
In the spectrum zone affected only by DCR and crosstalk, CNR becomes the crosstalk probability,

\begin{equation}
    \text{CT} = \frac{N_{\text{1.5pe}}}{N_{\text{0.5pe}}}.
\end{equation}

However, in the afterpulsing region, $N_{BLN}$ = $N_{\text{0.5pe}} + N_{AP}$, where $N_{AP}$ is the number of afterpulse events.

\begin{equation}
    \text{CNR} = \frac{N_{\text{1.5pe}}}{N_{\text{0.5pe}} + N_{AP}}.
\end{equation}
Then, the afterpulse probability can be estimated as,

\begin{equation}
\label{eq:ap}
    \text{AP} = \frac{N_{AP}}{N_{\text{0.5pe}}} = \frac{\text{CT}}{\text{CNR}}-1 .
\end{equation}

\begin{figure}[h!]
    \centering
    \includegraphics[width=0.47\linewidth]{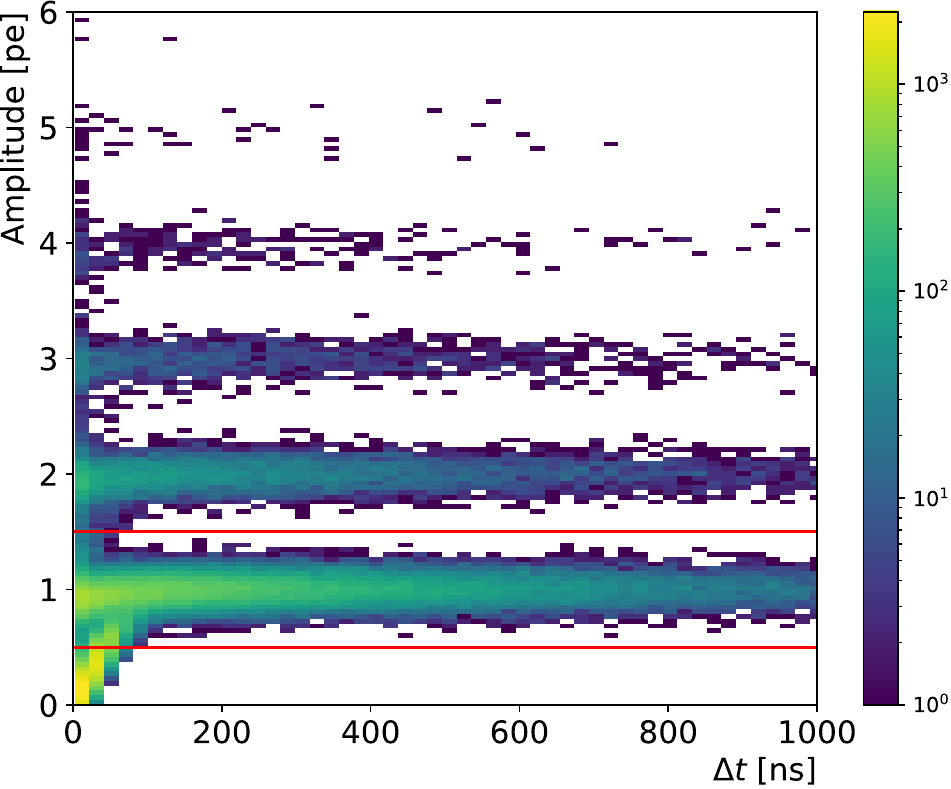}
    \includegraphics[width=0.52\linewidth]{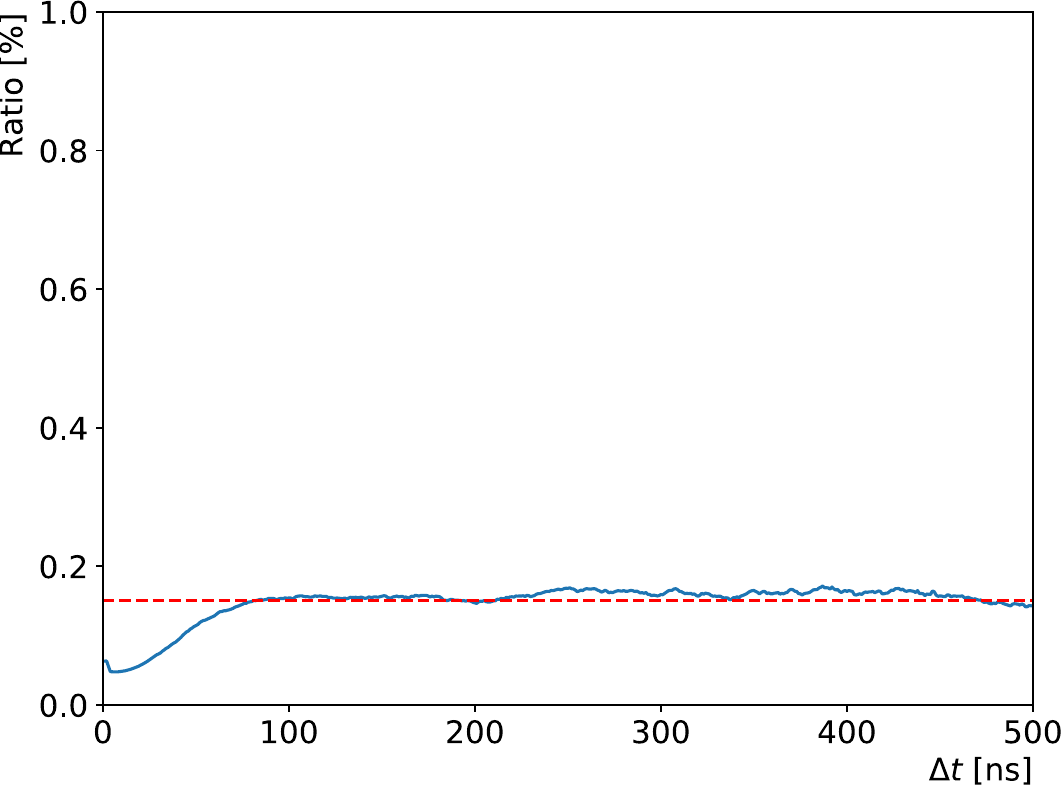}
    \caption{Amplitude vs.\ relative time spectrum coming out from AFBR-S4N66P024M SiPM simulations (left). CNR of an AFBR-S4N66P024M SiPM with crosstalk of 15$\%$ and afterpulsing of 60$\%$ (right). The red dashed line represents the simulated crosstalk, and the blue line represents the estimated CNR.}
    \label{fig:noise_sim}
\end{figure}

We carried out SiPM noise simulations to assess the reliability of the methodology \cite{PeaRodrguez2025}. Figure\,\ref{fig:noise_sim} (left) shows the simulated amplitude vs.\ relative time spectrum assuming pulse characteristics of an AFBR-S4N66P024M SiPM with 15$\%$ crosstalk and 60$\%$ afterpulsing. The derived correlated noise ratio (CNR, shown in \ref{fig:noise_sim} (right) as function of $\Delta t$) is equal to the input crosstalk probability in the region where the SiPM noise is not affected by afterpulsing ($\Delta t > 100$\,ns), but it decreases in the afterpulsing region.

\begin{figure}[h!]
    \centering
    \includegraphics[width=0.48\linewidth]{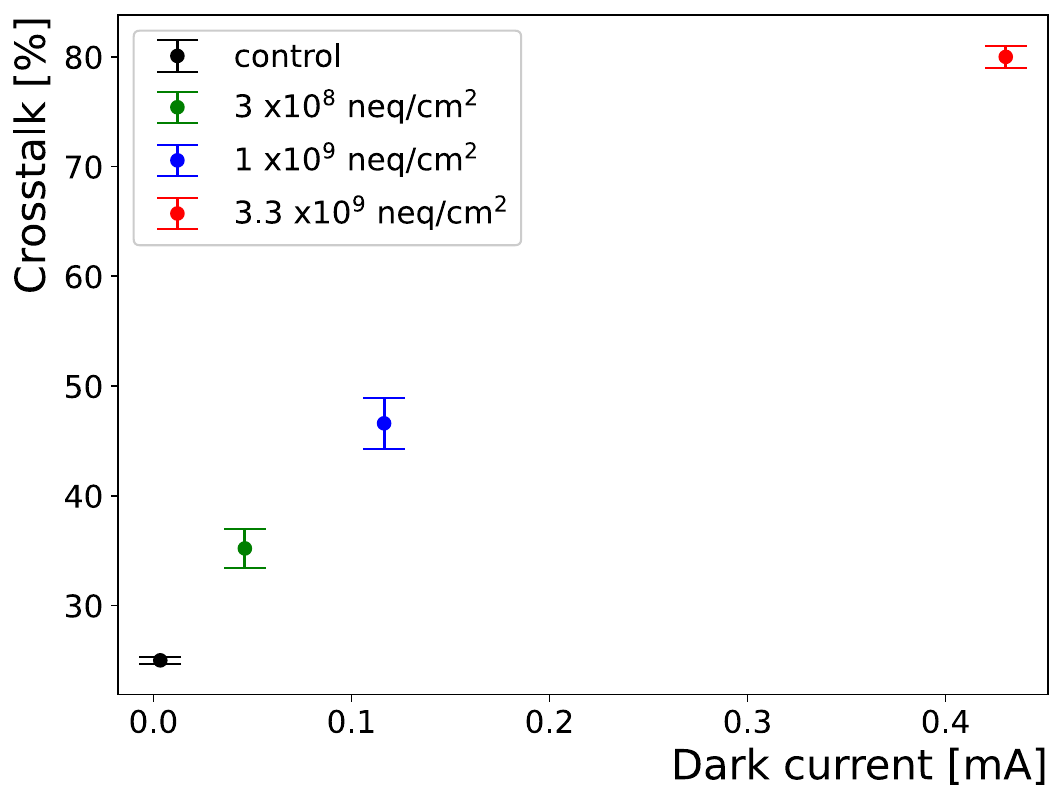}
    \includegraphics[width=0.48\linewidth]{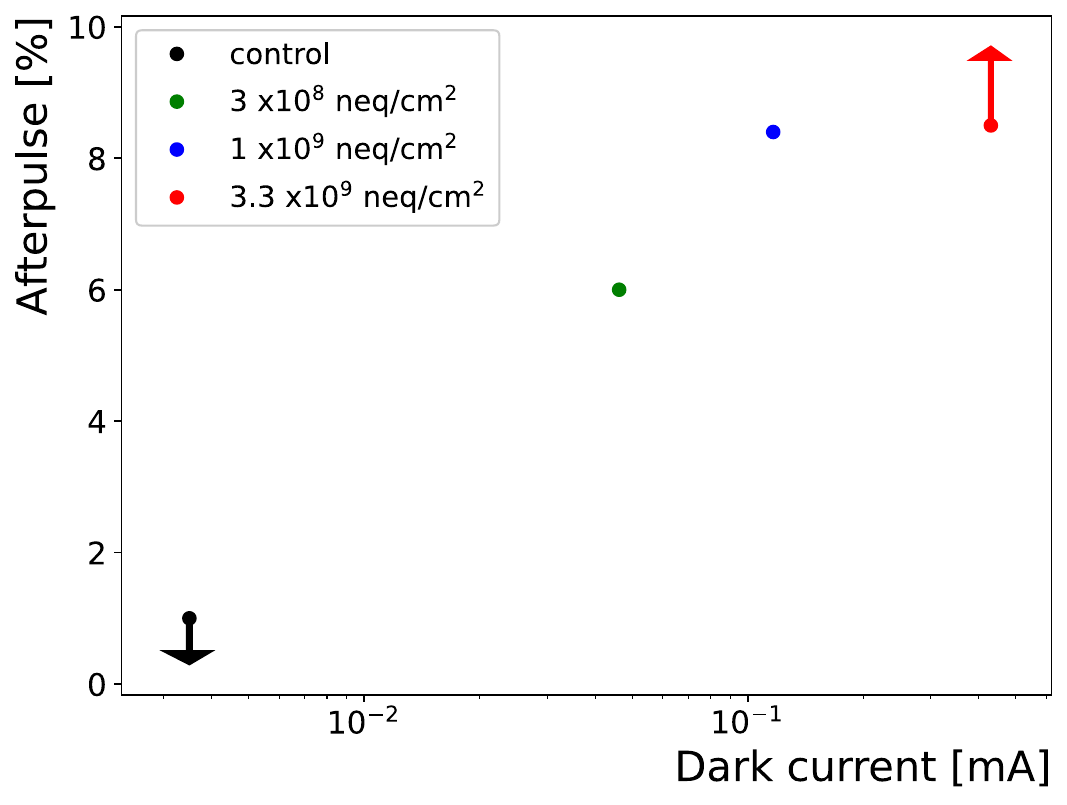}
    \caption{Crosstalk (left) and afterpulse (right) probability of AFBR-S4N66P024M SiPMs irradiated at $3 \times 10^8$\,n$_{\text{eq}}$/cm$^2$, $1 \times 10^9$\,n$_{\text{eq}}$/cm$^2$, and $3.3 \times 10^9$\,n$_{\text{eq}}$/cm$^2$.}
    \label{fig:corr_noise}
\end{figure}

Applying this methodology, we estimated the crosstalk and afterpulsing probability of AFBR-S4N66P024M SiPMs after neutron irradiation and annealing. Figure\,\ref{fig:corr_noise} shows the crosstalk and afterpulsing of the (non-irradiated) control sample and three irradiated samples ($3 \times 10^8$\,n$_{\text{eq}}$/cm$^2$, $1 \times 10^9$\,n$_{\text{eq}}$/cm$^2$, and $3.3 \times 10^9$\,n$_{\text{eq}}$/cm$^2$) annealed at 250\,$^{\circ}$C/30\,min. 

An increase in crosstalk could be caused by damage of trenches after irradiation allowing more photons to reach neighboring cells, changes in the refractive index of the SiPM window allowing back-reflection of photons reaching neighboring cells, or pileup of DCR events (fake crosstalk) \cite{Altamura2022}. In this case, the main factor that contributes to the increase in crosstalk shown in Fig.\,\ref{fig:corr_noise} (left) is the DCR pileup. We corroborated this through simulations that consecutive DCR signals can mimic a crosstalk event. Even when applying the differentiation algorithm, if the time resolution of the data acquisition system is insufficient, consecutive DCR events cannot be disentangled.

We observed an increase in the afterpulse probability with the radiation dose as shown in Fig.\,\ref{fig:corr_noise} (right). In the control sample, the estimated afterpulse probability rounds 1\,$\%$ and reaches $>8\,\%$ for an irradiation dose of $3.3 \times 10^9$\,n$_{\text{eq}}$/cm$^2$. These results clearly demonstrate that neutron radiation triggers trap creation in the silicon lattice, raising the afterpulse probability.

\subsection{Single-photon resolution}

Single-photon resolution is defined as the capability to differentiate the number of detected photons. Neutron irradiation affects the single-photon resolution of SiPMs. An increase in DCR and afterpulsing creates events with intermediate amplitudes between photon-equivalent levels. This process widens the amplitude distribution for single photons, resulting in a broader standard deviation. The photon resolution is given by:
\begin{equation}
    R_{ph} = \frac{\sigma_{pe}}{\Delta_{pe}} 100\%
\end{equation}
where $\sigma_{pe}$ is the standard deviation of a photon level distribution and $\Delta_{pe}$ is the difference between mean values of consecutive photon level distributions.

\begin{figure}[h!]
    \centering
    \includegraphics[width=0.9\linewidth]{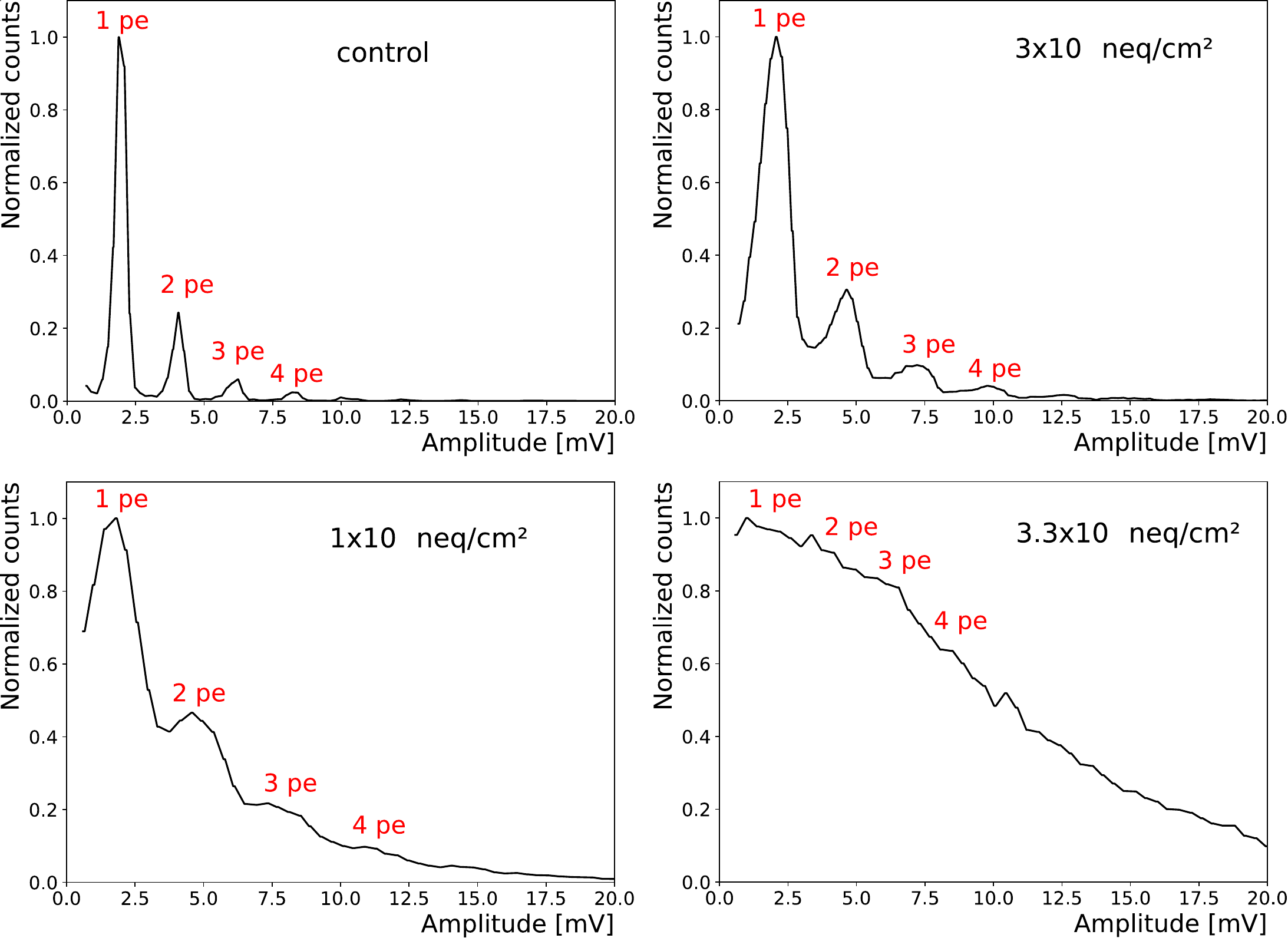}
    \caption{Normalized amplitude spectrum of the irradiated and annealed samples operating at 43\,V. The single-photon resolution degrades as the irradiation dose increases.}
    \label{fig:photon_resolution}
\end{figure}

Figure\,\ref{fig:photon_resolution} shows the amplitude spectrum of the AFBR-S4N66P024M SiPMs operated at 43\,V after irradiation and annealing (at 250\,$^{\circ}$C for 30\,min) operating at 43\,V. An evident degradation of the single-photon resolution is observed as the irradiation dose increases. In the control sample, peaks at the photon-level are well distinguishable resulting in a photon resolution $\sim 24$\,\%, while for the sample irradiated at $3 \times 10^8$\,n$_{\text{eq}}$/cm$^2$, inter-photon-level gaps are filled due to the widening of the individual photon-level distributions, degrading the resolution to $\sim 58$\,\%. The resolution of the sample irradiated at $1 \times 10^9$\,n$_{\text{eq}}$/cm$^2$ was $\sim 140$\,\% which means that the standard deviation of the single-photon peaks is larger than the separation between consecutive peaks. Estimating the photon resolution at higher irradiation doses becomes difficult because individual photon-levels vanish completely.

\subsection{Gain, capacitance and depletion zone}

The creation of vacancies, interstitials, dislocations, and clusters within the silicon lattice modifies the inner properties of APDs that comprise the SiPM. These changes are reflected in modifications to the doping concentration within the depletion region, the removal or creation of donors/acceptors, and increases or decreases in the negative and positive space charge. These processes affect characteristics such as gain and capacitance; however, the specific effect depends on the silicon structure of the SiPM APDs \cite{Kushpil2015,Garutti2019,Musienko2000}.

The SiPM gain is given by
\begin{equation}
    G = \frac{Q}{e},
\end{equation}
where $Q$ is the avalanche charge and $e$ is the electron charge ($1.6 \times 10^{-19}$\,C). The avalanche charge is deduced from the SiPM voltage signal $v(t)$ by
\begin{equation}
    Q = \frac{1}{A_vR}\int v(t)dt,
\end{equation}
where $R$ is the load resistance of the SiPM readout circuit and $A_v$ is the circuit amplification gain.

\begin{figure}[t]
    \centering
    \includegraphics[width=0.6\linewidth]{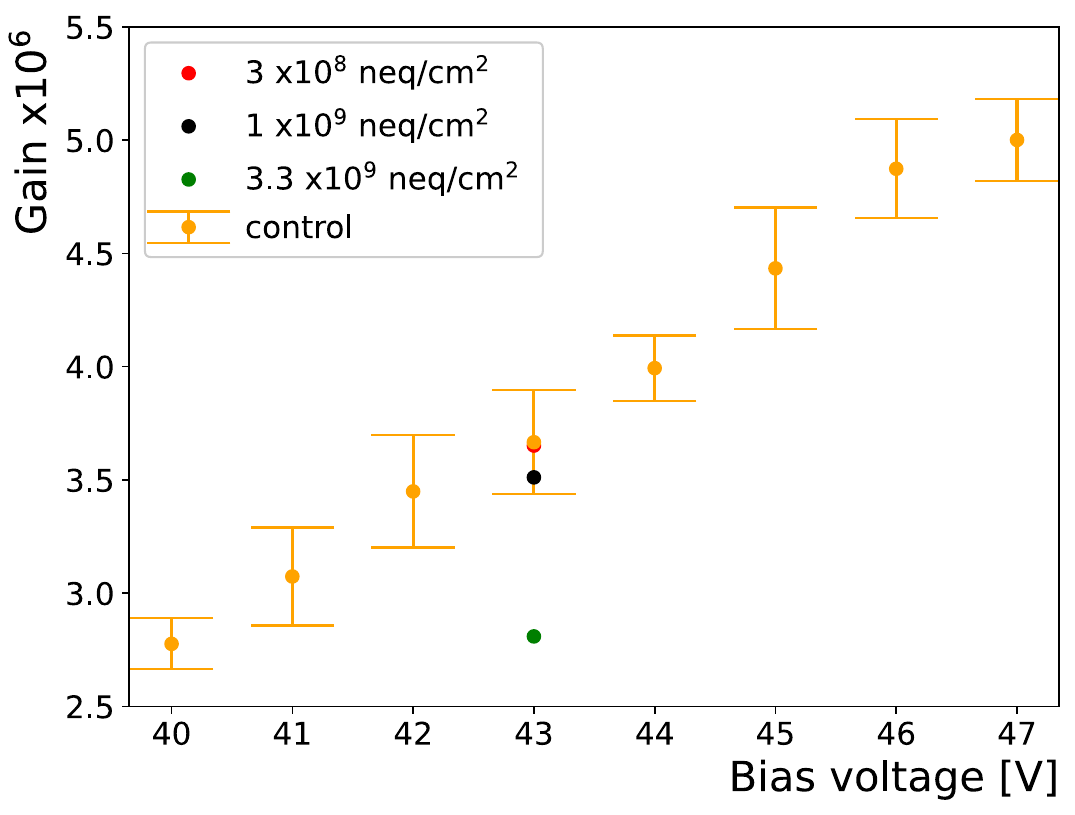}
    \caption{Gain of the AFBR-S4N66P024M APDs after neutron irradiation.}
    \label{fig:gain}
\end{figure}

We measured the gain of the non-irradiated AFBR-S4N66P024M reference sample as a function of the bias voltage (40~V to 47~V), and compared it to the gain of the irradiated samples at 43~V bias, as shown in Fig.\,\ref{fig:gain}. The variation in gain between the different sensors before irradiation is found to be rather small, justifying this comparison.
Thus, we conclude that the gain is only slightly affected for samples irradiated at $3 \times 10^8$\,n$_{\text{eq}}$/cm$^2$ and $1 \times 10^9$\,n$_{\text{eq}}$/cm$^2$ maintaining a value around $3.5 \times 10^6$ at 43\,V. However, at $3.3 \times 10^9$\,n$_{\text{eq}}$/cm$^2$, the gain decreases below $3 \times 10^6$.

\begin{figure}[t]
    \centering
    \includegraphics[width=0.49\linewidth]{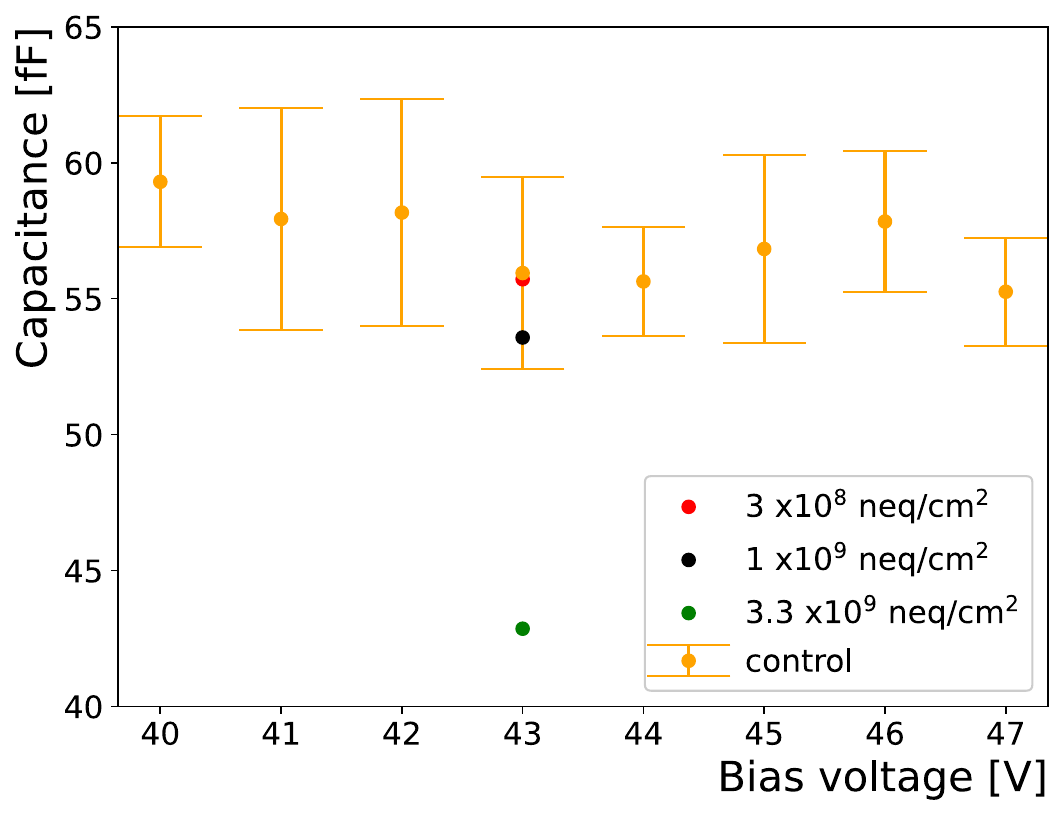}
    \includegraphics[width=0.49\linewidth]{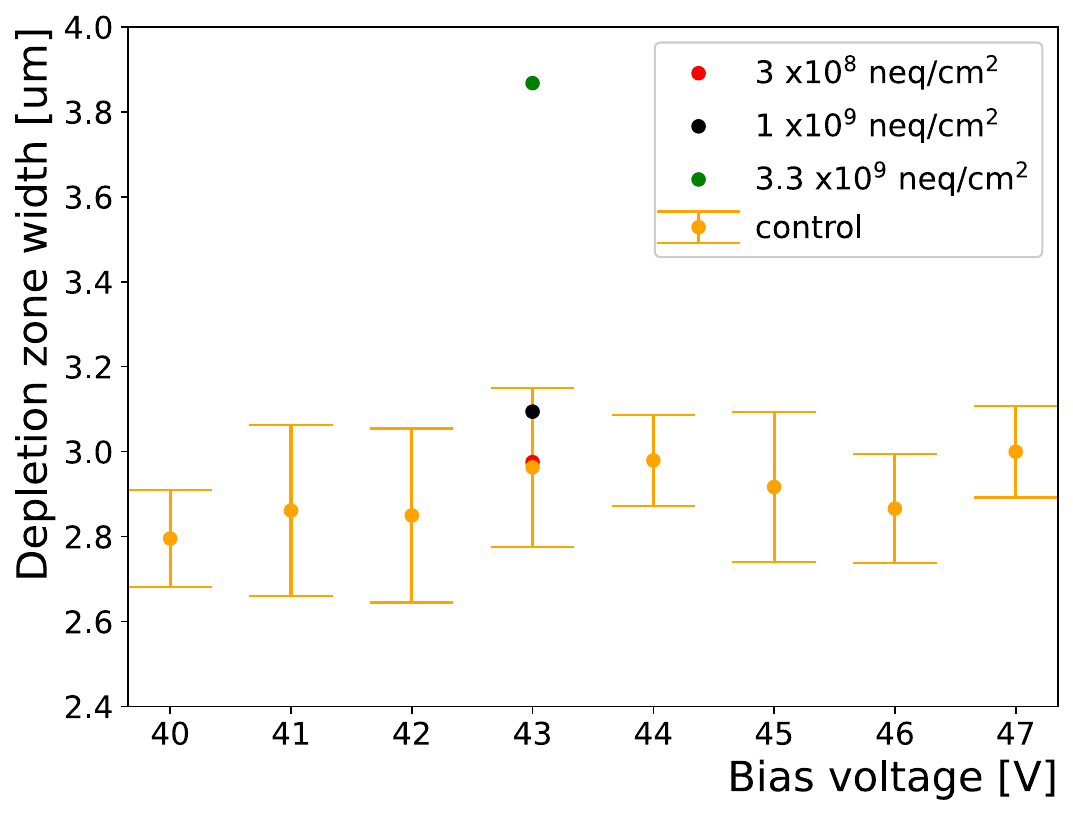}
    \caption{Capacitance (left) and depletion zone width (right) of the AFBR-S4N66P024M APDs after neutron irradiation.}
    \label{fig:capacitance}
\end{figure}

The capacitance of a single APD is proportional to the gain by
\begin{equation}
\label{eq:capcitance}
    G=\frac{C_d \left(V_{bias}-V_{bd}\right)}{e},
\end{equation}
where $V_{bd}$ and $C_d$ are the breakdown voltage and the APD junction capacitance respectively. Figure\,\ref{fig:capacitance} (left) shows the capacitance of the AFBR-S4N66P024M APDs before and after irradiation/annealing, operating with a constant breakdown voltage of 32.5\,V. The capacitance keeps constant ($\sim$57\,fF) and independent of the bias voltage before irradiation. However, the capacitance drops after neutron irradiation, reaching $\sim$42.5\,fF at $3.3 \times 10^9$\,n$_{\text{eq}}$/cm$^2$.

The width of the depletion zone of the APD is given by
\begin{equation}
    d=\frac{\epsilon_{0} \ \epsilon_{Si} \ \epsilon_{geom} \ A}{C_d},
\end{equation}
where $\epsilon_0$ is the vacuum permittivity ($8.854 \times 10^{-12}$\,ASV$^{-1}$m$^{-1}$),  $\epsilon_{Si} = 11.7$ relative permittivity of silicon, $\epsilon_{geom}$ is the corresponding geometric acceptance, and $A$ is the area of the APD depletion zone. 

Neutron irradiation causes the depletion zone in the AFBR-S4N66P024M APDs to broaden, increasing its size from $\sim2.8$\,\textmu m before irradiation to $\sim3.9$\,\textmu m after irradiation at $3.3 \times 10^9$\,n$_{\text{eq}}$/cm$^2$ as shown in Fig.\,\ref{fig:capacitance} (right). This variation in the depletion zone affects the SiPM PDE. 


\subsection{Annealing optical effects}

\begin{figure}[t]
    \centering
    \includegraphics[width=1\linewidth]{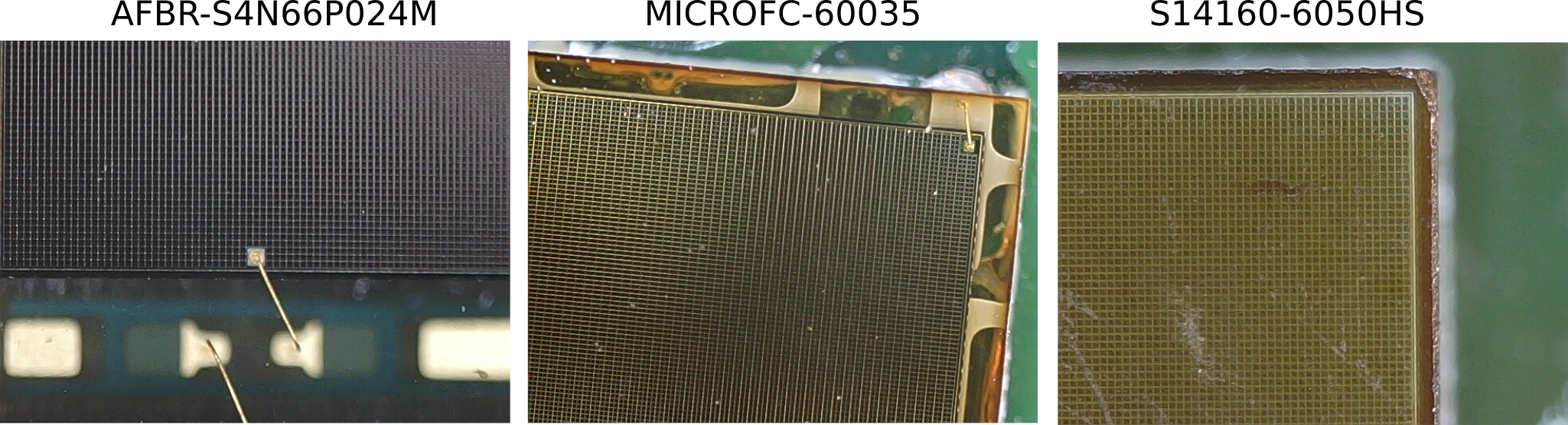}
    \caption{Annealing effects in the SiPMs entrance window. Microscopic view of the photon entrance window of the AFBR-S4N66P024M, MICROFC-60035, and S14160-6050HS SiPMs.}
    \label{fig:window}
\end{figure}

Due to the degradation of the entrance window, annealing can affect the optical properties of SiPMs, as discussed in \cite{Calvi2019}. The extend of the degradation depends on the material of the window. Figure\,\ref{fig:window} shows a microscopic view of the entrance window of the AFBR-S4N66P024M, MICROFC-60035, and S14160-6050HS SiPMs after annealing at 250\,$^{\circ}$C for 30\,min. Changes in the window color are observed for the MICROFC-60035 and S14160-6050HS (silicon resin) samples. The epoxy window of the AFBR-S4N66P024M remains transparent after annealing.

\section*{Conclusions}

In this study, we conducted a throughout performance analysis of three different SiPMs (AFBR-S4N66P024M, MICROFC-60035, and S14160-6050HS) following neutron irradiation and electrical annealing as a recovery method. The neutron radiation dose ranged from $3 \times 10^8$\,n$_{\text{eq}}$/cm$^2$ to  $1 \times 10^{11}$\,n$_{\text{eq}}$/cm$^2$. We observed that the dark current increased by a factor of 200 at a dose of $3 \times 10^8$\,n$_{\text{eq}}$/cm$^2$, while at  $1 \times 10^{11}$\,n$_{\text{eq}}$/cm$^2$ the factor was 2000. This increase in dark current is due to point and cluster defects caused by neutron interactions in the silicon lattice. We implemented an electrical annealing system to recover the SiPM performance. The annealing temperatures of 160\,$^{\circ}$C and 250\,$^{\circ}$C were controlled by applying a constant forward current of 290-420\,mA through the SiPM. After annealing at 250\,$^{\circ}$C for 30\,minutes, the unannealed damage R at $3 \times 10^8$\,n$_{\text{eq}}$/cm$^2$ was 1.35\,\%, while for the sample irradiated at $1 \times 10^{11}$\,n$_{\text{eq}}$/cm$^2$ R was 52.6\,\%.

Neutron irradiation affects the noise components of the SiPM: DCR, crosstalk, and afterpulsing. We found that the dark noise increases exponentially with the neutron dose. The AFBR-S4N66P024M control sample, which was operated at 43\,V, had a DCR of 3.8\,MHz ($I_{dc}$ = 7.35\,\textmu A). At a dose of $3 \times 10^8$\,n$_{\text{eq}}$/cm$^2$, it increased to 26\,MHz ($I_{dc}$ = 52\,\textmu A), and at $3.3 \times 10^9$\,n$_{\text{eq}}$/cm$^2$ it reached 51.5\,MHz ($I_{dc}$ = 122\,\textmu A). Neutron irradiation introduces nonlinearities in the dark current and DCR relationship. Those come from afterpulsing created by traps in the silicon lattice and by the pileup of dark count events (fake crosstalk). In this case, the afterpulsing increased from 1$\%$ (control) to $> 8\%$ ($3.3 \times 10^9$\,n$_{\text{eq}}$/cm$^2$).

An unintended consequence of the increase in dark noise was the degradation of the SiPM single-photon resolution. We observed that the single-photon resolution decreases from 24\,\% in the control sample to 140\,\% at $3.3 \times 10^9$\,n$_{\text{eq}}$/cm$^2$. Changes in the silicon doping and the depletion region were also found. These changes affect the SiPM gain and capacitance. The gain was 3.5$\times 10^6$ (57\,fF) at a dose of $3 \times 10^8$\,n$_{\text{eq}}$/cm$^2$, which decreased to 3$\times 10^6$ at $3.3 \times 10^9$\,n$_{\text{eq}}$/cm$^2$ (42.5\,fF).

Understanding the behavior of SiPMs in high radiation environments helps design robust triggering and readout systems, as well as implement actions to compensate for radiation effects. Planning of annealing campaigns after each beam-time can recover the general performance of the SiPM and extend its lifetime before replacement. Specific actions, such as cooling the SiPM or tuning the bias voltage, can compensate for radiation effects on dark noise and gain, respectively. 

\section*{Acknowledgments}

This work is supported by Netzwerke 2021, an initiative of the Ministry of Culture and Science of the State of North Rhine Westphalia.

\bibliographystyle{elsarticle-num} 
\bibliography{SiPM_fram}






\end{document}